\documentclass[%
 reprint,
 amsmath,amssymb,
 aps,
pra,
]{revtex4-2}

\usepackage{graphicx}% Include figure files
\usepackage{dcolumn}% Align table columns on decimal point
\usepackage{bm}% bold math
\usepackage{physics}
\newcommand{\lsup}{\bar\lambda^{\scriptscriptstyle W}_1}
\begin{document}

\preprint{APS/123-QED}

\title{Construction of genuinely multipartite entangled subspaces\\ and the associated bounds  on entanglement measures for mixed states}% Force line breaks with \\

\author{K. V. Antipin}
 \email{kv.antipin@physics.msu.ru}
\affiliation{%
 Lomonosov Moscow State University
}%


\date{\today}% It is always \today, today,

\begin{abstract}
Genuine entanglement is the strongest form of multipartite entanglement. Genuinely entangled pure states contain entanglement in every bipartition and as such can be regarded as a valuable resource in the protocols of quantum information processing. A recent direction of research is the construction of genuinely entangled subspaces --- the class of subspaces consisting entirely of genuinely multipartite entangled pure states. In this paper we present several methods of construction of such subspaces including those of maximal possible dimension. The approach is based on the correspondence between bipartite entangled subspaces and quantum channels of a certain type. The examples include maximal subspaces for systems  of three qubits, four qubits,  three qutrits. We also provide lower bounds on  two entanglement measures for mixed states, the concurrence and the convex-roof extended negativity, which are directly connected with the projection on genuinely entangled subspaces.
%\begin{description}
%use the optional argument of the \verb+\item+ command to give the category of %each item. 
%\end{description}
\end{abstract}

\maketitle

%\tableofcontents

\section{Introduction}
Quantum entanglement is an incredibly rich phenomenon relevant to a wide range of fields from condensed matter physics to quantum information science to particle physics. In quantum information theory entanglement is regarded as a resource for the tasks of quantum communication and computation~\cite{Wilde13,Preskill}.

The structure of multipartite entanglement is far more complex than that in bipartite systems. There are various inequivalent entanglement classes~\cite{DVC00,VDMV02}. There are also such peculiar properties  as  monogamy relations~\cite{TerhalMon04,CKW00,KW04}  exhibited by some correlations of particles.

Genuine multipartite entanglement~(GME)~\cite{GME99} is an extreme form of the described multipartite phenomenon. Pure GME states are entangled with respect to any bipartition of a compound system, the property that has recently found applications in the protocols of quantum information processing~(see, e.~g., \cite{YeCh06,MP08,MEO18}).

An interesting direction of research is construction and characterization of genuinely entangled subspaces~(GESs) --- subspaces consisting entirely of genuinely multipartite entangled states. To our knowledge, this concept was first introduced in Ref.~\cite{DemAugWit18}, where some approaches to construction of GESs were also proposed. GESs can be interesting in connection with generating mixed GME  states --- any state with support on a GES is genuinely entangled. It is believed that GESs will be useful in quantum error correction~\cite{HuGra20} and quantum cryptography~\cite{SheSrik19}.

The research into the new methods of constructing GESs was continued in Refs.~\cite{DemAugQut19,WaChenZhGu19,AgHalBa19,DemAug20}. In Refs.~\cite{DemAugQut19,WaChenZhGu19,AgHalBa19} the construction was based mainly on the unextendible product bases~(UPB) method~\cite{BDMSST99,DMSST03} and its variations. The approach in Ref.~\cite{DemAug20} was aimed at construction of GESs of maximal possible dimension and relied on the characterization of some bipartite entangled subspaces.

The motivation of the present paper is to develop an alternative approach to the construction of GES which is not based on UPB. At the center of our considerations is the correspondence between quantum channels and subspaces of a tensor product Hilbert space~(see Section~\ref{sec::prel}). In many cases such technique allows us to obtain  families of GESs with simple parameterization, including those of maximal possible dimension. Another aspect is detecting entanglement of mixed states and estimating entanglement measures for them. It is known that there is an entanglement witness associated with the projection on an entangled bipartite subspace~\cite{Sar08}. The extension of this witness to the GME case was obtained in Ref~\cite{DemAugWit18}. In connection with the~(bipartite) witness also lower bounds on some entanglement measures for bipartite mixed states can be derived~\cite{KVAnt20}. We aim to extend these bounds to the GME case and use the examples of  GESs we construct to illustrate their application. It was shown that computation of the convex-roof entanglement measures is NP-complete~\cite{Huang14}, so, in genneral, efficient algorithms or even closed mathematical expressions for them are impossible unless $\mathrm P = \mathrm{NP}$. In this regard such bounds play an important role in the theory of entanglement.

The paper is organized as follows. In Section~\ref{sec::prel} we provide the necessary definitions and theoretical background. The methods of constructing GESs and their application to concrete three- and four-partite systems are presented in Section~\ref{sec::const}. In Section~\ref{sec::EM} we provide estimation of entanglement of some constructed subspaces and talk about lower bounds on entanglement measures for mixed states connected with overlap  with a given GES. We conclude in Section~\ref{sec::conc} and discuss possible developments of the present line of research.

\section{Preliminaries}\label{sec::prel}
\subsection{Genuine multipartite entanglement}

Consider $n$-partite quantum states in the finite dimensional tensor product Hilbert space $H_1\otimes\ldots\otimes H_n$. A pure $n$-partite state $\ket{\psi}$ is called \emph{fully separable} if it can be written as a tensor product of states for every subsystem,~i.~e.
\begin{equation}
    \ket{\psi}=\ket{\phi}_1\otimes\ldots\otimes\ket{\phi}_n.
\end{equation}
States that are not fully separable are said to be \emph{entangled}.

A pure $n$-partite state $\ket{\psi}$ is called \emph{biseparable} if it can be written as a tensor product
\begin{equation}
    \ket{\psi} = \ket{\psi}_A\otimes\ket{\psi}_{\bar{A}}
\end{equation}
with respect to some bipartition $A|\bar A$ where $A$ denotes a particular subset of subsystems and $\bar A$ denotes its complement. A multipartite pure state is called \emph{genuinely multipartite entanged} if it is not biseparable with respect to any bipartition.

There is generalization of these concepts to mixed states. A mixed  multipartite state is called \emph{biseparable} if it can be decomposed into a convex sum of biseparable pure states~(note that different terms of the sum can be biseparable with respect to different bipartitions). Otherwise it is called \emph{genuinely multipartite entanged}.

A subspace of a multipartite Hilbert space consisting entirely of entangled pure states is called \emph{completely entangled}~(CES). The examples of CESs are known~\cite{Parth04,Bhat06,CMW08}. The natural generalization of this notion is \emph{genuinely multipartite entangled subspaces}~(GESs) -- those composed entirely of genuinely entangled pure states.

\subsection{Entanglement measures of states and subspaces}
For the purpose of quantifying entanglement  many entanglement measures were introduced, initially for bipartite systems~\cite{PlenVir07,Hor09,GuhnToth09}. Among them -- the concurrence~\cite{BenVinSmol96,Woot98,BDH02}, the negativity~\cite{VidWer02,LCOK03} and the geometric measure of entanglement~\cite{BarnLin01,WeiGold03}.

The \emph{concurrence} of a pure bipartite state $\psi$ is defined by
\begin{equation}\label{purconc}
    C(\psi) = \sqrt{2\left(1-\Tr{\rho_A^2}\right)},
\end{equation}
where $\rho=\dyad{\psi}$ and $\rho_A = \mathrm{Tr}_B\{\rho\}$.

For mixed states $\rho$ the concurrence is given by the convex roof construction, the minimum average concurrence taken over all ensemble decompositions of $\rho$:
\begin{equation}\label{concconv}
    C(\rho) = \min_{\{(p_j,\,\psi_j)\}}\,\sum_j\,p_j\,C(\psi_j).
\end{equation}

The \emph{negativity} of $\rho$ is defined as 
\begin{equation}\label{negdef}
    N(\rho) = \frac12(\norm{\rho^{T_B}}_1 - 1),
\end{equation}
where $\rho^{T_B}$ is the partial transpose of $\rho$ with respect to party $B$, and $\norm{A}_1 = \Tr{\sqrt{A^{\dagger}A}}$ is the trace norm of $A$.

From the definition of $N$ it is seen that entanglement of states with a positive partial transpose~(PPT states) is \emph{not detected} by this measure.

The \emph{convex-roof extended negativity}~(CREN) is given by
\begin{equation}\label{cren}
    N^{\mathrm{CREN}}(\rho) = \min_{\{(p_j,\,\psi_j)\}}\,\sum_j\,p_j\,N(\psi_j).
\end{equation}

The two measures defined via convex roof are able to detect \emph{all} entangled states.

Given the Schmidt decomposition $\ket{\psi} = \sum_i\,\sqrt{\lambda_i}\ket{i}\otimes\ket{i}$ of a bipartite pure state the geometric measure of entanglement is defined by
\begin{equation}\label{geom}
    G(\psi) = 1 - \max_i\{\lambda_i\}.
\end{equation}
This measure is also extended to mixed states by the convex roof construction.

Bipartite entanglement measures can be generalized to measures of genuine multipartite entanglement~\cite{Guhnetall20}. Given a bipartite entanglement measure $E$, the corresponding GME measure for a pure multipartite state $\ket{\psi}$ is defined as
\begin{equation}\label{geomGME}
    E_{GME}(\psi) = \min_A E_A(\psi),
\end{equation}
where the minimum is taken over all possible bipartitions $A|\bar A$ of a multipartite system and $E_A$ denotes the bipartite entanglement measure with respect to bipartition $A|\bar A$. The measure $E_{GME}$ is extended to mixed states by the convex roof construction
\begin{equation}\label{cfGME}
    E_{GME}(\rho) = \min_{\{(p_j,\,\psi_j)\}}\,\sum_j\,p_j\,E_{GME}(\psi_j),
\end{equation}
where, as usual, the minimization runs over all possible ensemble decompositions $\rho = \sum_j p_j\dyad{\psi_j}$.

There are several ways of quantifying entanglement of a subspace. In the present paper we will use the approach suggested in Refs.~\cite{GourWall07},\cite{DemAugQut19}: the entanglement of a subspace $W$ measured by $E$ is defined by
\begin{equation}
   E(W) =  \min_{\ket{\psi}\in W}E(\psi).
\end{equation}
In place of $E$ here we will use the geometric measure of genuine entanglement, $G_{GME}$, which can be calculated via eqs.~(\ref{geom}) and (\ref{geomGME}).

\subsection{Quantum channels and entangled subspaces}

Let $H$ be a finite dimensional Hilbert space, and $\mathcal B(H)$~-- a collection of linear operators on $H$. Given two finite dimensional Hilbert spaces $H_A$ and $H_B$,  a \emph{quantum channel} is a linear, completely positive and trace-preserving map between $\mathcal B(H_A)$ and $\mathcal B(H_B)$ \cite{Wilde13}. By definition, a quantum channel maps density operators to density operators.

Action of a quantum channel $\mathrm\Phi\colon\,\mathcal B(H_A)\rightarrow\mathcal B(H_B)$ on an operator $C \in\mathcal B(H_A) $, by Choi's theorem\cite{Choi75}, can be represented as
\begin{equation}\label{Kraus}
    \mathrm\Phi(C) = \sum_{i=1}^N\,K_i C K_i^{\dagger},
\end{equation}
where $N\leqslant\mathrm{dim}(H_A)\,\mathrm{dim}(H_B)$, and the operators $K_i\colon\,H_A\rightarrow H_B$ satisfy the trace-preserving property
\begin{equation}\label{TrP}
    \sum_{i=1}^N\, K_i^{\dagger}K_i = I,
\end{equation}
where $I$ -- the identity operator, which in this case acts on $H_A$. Representation~(\ref{Kraus}) is often referred to as the Kraus decomposition of a quantum channel with Kraus operators $\{K_i\}$.

There is a one-to-one correspondence between channels and linear subspaces of composite Hilbert spaces\cite{AubSz17}. Given a set of three Hilbert spaces $(H_A,\, H_B,\, H_C)$, a subspace $W$ of the composite space $H_B\otimes H_C$, and an isometry $V\,\colon\, H_A\rightarrow H_B\otimes H_C$ whose range is $W$, the corresponding quantum channel $\mathrm\Phi\colon\,\mathcal B(H_A)\rightarrow\mathcal B(H_B)$ can be defined by
\begin{equation}\label{IsoRep}
    \mathrm\Phi(\rho) = \mathrm{Tr}_{H_C} (V\rho V^{\dagger}).
\end{equation}
Conversely, due to Stinespring's dilation theorem~\cite{Stine55}, any quantum channel $\mathrm\Phi\colon\,\mathcal B(H_A)\rightarrow\mathcal B(H_B)$ can be represented by eq.~(\ref{IsoRep}) for some subspace $W\subset H_{B}\otimes H_{C}$.

Given some Kraus representation~(\ref{Kraus}) of a quantum channel $\mathrm\Phi\colon\,\mathcal B(H_A)\rightarrow\mathcal B(H_B)$, the isometry~$V\,\colon\, H_A\rightarrow H_B\otimes H_C$ can be defined by the action on a vector state $\ket{\psi}\in H_A$ in the following way:
\begin{equation}\label{IsoK}
    V\ket{\psi} = \sum_{i=1}^N\,K_i\ket{\psi}\otimes\ket{i},
\end{equation}
where $\{\ket{i}\}$ -- an orthonormal basis in $H_C$, and $\mathrm{dim}(H_C) = N$. Choosing some orthonormal basis $\{{\ket{\mu}}\}$ in $H_A$, via eq.~(\ref{IsoK}) we obtain an orthonormal system of vectors $\{V\ket{\mu}\}$ spanning the subspace $W\subset H_{B}\otimes H_{C}$.

\begin{figure}[b]
\includegraphics[scale=0.35]{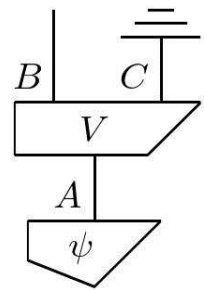}% Here is how to import EPS art
\caption{\label{fig:channel} A representation of quantum channel $\mathrm\Phi$ acting on a pure state $\ket{\phi}\in H_A$: at first, isometry $V$ takes the state  to $H_B\otimes H_C$, then subsystem $C$ is traced out. The result is some density operator acting on $H_B$.}
\end{figure}

Fig.~\ref{fig:channel} provides the diagrammatic representation of eq.~(\ref{IsoRep}). We adopt tensor diagram notation for quantum processes from Ref.~\cite{CoeKis17}. Refs.~\cite{BiamonteEtAll15}, \cite{Biamonte19} are also excellent introductions into diagrammatic reasoning in quantum information theory.

The entanglement of the subspace $W\subset H_{B}\otimes H_{C}$ depends on the output characteristics of the corresponding quantum channel such as the maximal output norm \cite{Holev00}. Let $\mathcal D(H)$ denote the set of density operators in $\mathcal B(H)$. \emph{The maximal output norm} of a channel $\mathrm\Phi\colon\,\mathcal B(H_A)\rightarrow\mathcal B(H_B)$  is defined by
\begin{equation}\label{OutNorm}
    \nu_p(\mathrm\Phi) = \sup_{\rho\in\mathcal D(H_A)}\norm{\mathrm\Phi(\rho)}_p,\,\,p>1,
\end{equation}
where $\norm{\rho}_p = (\mathrm{Tr}(\abs{\rho}^p))^{1/p}$ is the standard $p$-norm. Since $p$-norm is convex, it will take its maximum on the extremal states, and the supremum in eq.~(\ref{OutNorm}) can be taken over pure input states. We will mostly use the Frobenius norm~($p = 2$) and refer to 
$(\nu_2(\mathrm\Phi))^2$ as the \emph{output purity} of a channel $\mathrm\Phi$.

Obviously, a subspace $W\subset H_{B}\otimes H_{C}$ is completely entangled if and only if the output purity of the corresponding quantum channel is strictly less than one. In this case the isometry $V$ will transform the vector states of $H_A$ into bipartite entangled states of the subspace $W\subset H_{B}\otimes H_{C}$, and there will not be separable states in $W$. Therefore, we can construct bipartite completely entangled subspaces finding examples of quantum channels with output purity less than one along with the corresponding isometries. Choosing some orthonormal basis $\{\ket{i}\}$ in $H_A$, we obtain the orthonormal basis $\{V\ket{i}\}$ of the entangled subspace $W$ via the isometry $V$. As we will see, the most interesting cases,  examples of entangled subspaces of maximal dimension, can be obtained by constructing dimension changing channels $\mathrm\Phi\colon\,\mathcal B(H_A)\rightarrow\mathcal B(H_B)$ with $\mathrm{dim}(H_A)\ne\mathrm{dim}(H_B)$. Furthermore, such channels can be used in construction of genuinely entangled  subspaces.

\begin{figure}[b]
\includegraphics[scale=0.35]{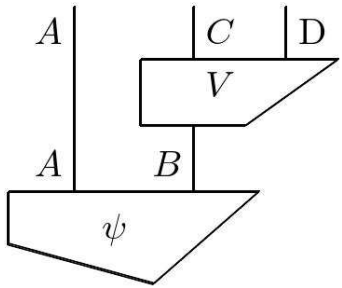}% Here is how to import EPS art
\caption{\label{fig:iso23} An isometry $V$ acting on  subsystem $B$ of  a pure bipartite entangled state $\ket{\psi_{AB}}$. The resulting pure state is tripartite, with subsystems $A$, $C$, $D$.}
\end{figure}

\section{\label{sec::const}Construction of entangled subspaces}

In this section we describe several methods of constructing completely and genuinely entangled subspaces with the use of quantum channels. We begin with the construction of  genuinely entangled subspaces from entangled subspaces of a bipartite Hilbert space.

\subsection{\label{sub::constch}GES from bipartite CES}

This method was inspired by tensor diagrams. Given a pure bipartite entangled state $\ket{\psi}_{AB}$, we can act on one of its two parties, $B$, with an isometry $V\,\colon\, H_B\rightarrow H_C\otimes H_D$ corresponding to some quantum channel $\mathrm\Phi\colon\,\mathcal B(H_B)\rightarrow\mathcal B(H_C)$, as it is illustrated on Fig.~\ref{fig:iso23}. When will the resulting state be entangled for any bipartite cut? It is convenient to consider the initial state $\ket{\psi}_{AB}$ in the Schmidt form:
\begin{equation}
    \ket{\psi}_{AB} = \sum_i\,\sqrt{p_i}\,\ket{i}_A\otimes\ket{i'}_B,
\end{equation}
where there are at least two nonzero Schmidt coefficients $\sqrt{p_i}$.
The resulting state
\begin{eqnarray}
    \ket{\chi}_{ACD} &=& (I\otimes V)\ket{\psi}_{AB}\nonumber\\
     & &= \sum_i\,\sqrt{p_i}\,\ket{i}_A\otimes V\ket{i'}_B
\end{eqnarray}
is surely entangled across bipartition $A|CD$: tracing out, for example, subsystem $CD$, we obtain a mixed state on $A$:
\begin{equation}\label{CESbi}
    \mathrm{Tr}_{CD}\left(\dyad{\chi}_{ACD}\right) = \sum_i\,p_i \dyad{i}_A.
\end{equation}

Now consider bipartition $C|AD$. Tracing out subsystem $A$, then $D$, we obtain:
\begin{eqnarray}\label{chbi}
\mathrm{Tr}_{AD}\left(\dyad{\chi}_{ACD}\right) &&\nonumber\\
&&=\sum_i\,p_i\mathrm{Tr}_D(V\dyad{i'}_B V^{\dagger})\nonumber\\
&&=\sum_i\,p_i\mathrm{\Phi}(\dyad{i'}_B)\label{partC},
\end{eqnarray}
where the second equality is due to eq.~(\ref{IsoRep}). For an arbitrary channel $\mathrm{\Phi}$ this state may be pure -- for example, we can choose a channel mapping all states to some pure state $\dyad{\phi}_C$. In order to guarantee mixedness, we can choose a channel with \emph{output purity less than one} -- in this case each term in eq.~(\ref{partC}) will be a mixed state, and the resulting state will be a convex combination of such mixed density operators. Entanglement across bipartition $D|AC$ is analyzed similarly. Therefore, we've established that genuinely entangled tripartite states can be obtained from entangled bipartite ones by applying to  one of the parties the isometry corresponding to a quantum channel with output purity less than 1. 

Now, applying such an isometry to each state in a completely entangled subspace of a bipartite Hilbert space, we obtain a genuinely entangled subspace of a tripartite Hilbert space, and, since isometry preserves inner products, the orthonormal system of vectors spanning the latter is obtained by action of $V$ on the orthonormal system spanning the former.

As an illustration of this approach we construct a four-dimensional genuinely entangled $3\otimes3\otimes3$ subspace from a completely entangled $3\otimes3$ subspace of the same dimension. Here the bipartite $3\otimes3$ subspace is taken to be spanned by the following orthonormal system of vectors:

\begin{subequations}\label{33ex}
\begin{equation}\label{33a}
    \ket{\psi_1} = \sqrt{\lambda_1} \ket{0}\otimes\ket{0} + \sqrt{1-\lambda_1}\ket{2}\otimes\ket{2},
\end{equation}
\begin{equation}
    \ket{\psi_2} = \sqrt{\lambda_2} \ket{1}\otimes\ket{0} + \sqrt{1-\lambda_2}\ket{0}\otimes\ket{1},
\end{equation}
\begin{equation}
    \ket{\psi_3} = \sqrt{\lambda_3} \ket{2}\otimes\ket{0} + \sqrt{1-\lambda_3}\ket{1}\otimes\ket{2},
\end{equation}
\begin{equation}\label{33d}
    \ket{\psi_4} = \sqrt{\lambda_4} \ket{2}\otimes\ket{1} + \sqrt{1-\lambda_4}\ket{0}\otimes\ket{2},
\end{equation}
\end{subequations}
where $0 < \lambda_i < 1,\:i = 1,\,2,\,3,\,4$.

In fact, eq.~(\ref{33ex}) determines a family of completely entangled subspaces parameterized by $\{\lambda_i\}$. They can be constructed with the use of specific dimension changing quantum channels~(see Appendix~\ref{app:33}).

Now, given a concrete $3\otimes3$ entangled subspace, we choose an isometry to act with on subsystem $B$, the second qutrit. As was shown above, the range of the isometry should be a completely entangled subspace, or, equivalently, the corresponding quantum channel should have output purity less than one. Such a role can play, for example,  the isometry $V\,\colon\, H_B\rightarrow H_C\otimes H_D$ which maps orthonormal basis in $H_B$ to vectors spanning the antisymmetric subspace of $H_C\otimes H_D$: 
\begin{subequations}
\begin{equation}
    V\ket{0} = \frac1{\sqrt2}(\ket{0}\otimes\ket{1} - \ket{1}\otimes\ket{0}),
\end{equation}
\begin{equation}
    V\ket{1} = \frac1{\sqrt2}(\ket{0}\otimes\ket{2} - \ket{2}\otimes\ket{0}),
\end{equation}
\begin{equation}
    V\ket{2} = \frac1{\sqrt2}(\ket{1}\otimes\ket{2} - \ket{2}\otimes\ket{1}).
\end{equation}
\end{subequations}
Acting with $(I\otimes V)$ on the basis~(\ref{33a}-\ref{33d}), we obtain the orthonormal system of vectors spanning  a four-dimensional genuinely entangled subspace:
\begin{subequations}
\begin{eqnarray}
\ket{\phi_1} = \sqrt{\frac{\lambda_1}2}(\ket{0}\otimes\ket{0}\otimes\ket{1} - \ket{0}\otimes\ket{1}\otimes\ket{0})\nonumber\\
+ \sqrt{\frac{1 -\lambda_1}2}(\ket{2}\otimes\ket{1}\otimes\ket{2} - \ket{2}\otimes\ket{2}\otimes\ket{1}),
\end{eqnarray}
\begin{eqnarray}
\ket{\phi_2} = \sqrt{\frac{\lambda_2}2}(\ket{1}\otimes\ket{0}\otimes\ket{1} - \ket{1}\otimes\ket{1}\otimes\ket{0})\nonumber\\
+ \sqrt{\frac{1 -\lambda_2}2}(\ket{0}\otimes\ket{0}\otimes\ket{2} - \ket{0}\otimes\ket{2}\otimes\ket{0}),
\end{eqnarray}
\begin{eqnarray}
\ket{\phi_3} = \sqrt{\frac{\lambda_3}2}(\ket{2}\otimes\ket{0}\otimes\ket{1} - \ket{2}\otimes\ket{1}\otimes\ket{0})\nonumber\\
+ \sqrt{\frac{1 -\lambda_3}2}(\ket{1}\otimes\ket{1}\otimes\ket{2} - \ket{1}\otimes\ket{2}\otimes\ket{1}),
\end{eqnarray}
\begin{eqnarray}
\ket{\phi_4} = \sqrt{\frac{\lambda_4}2}(\ket{2}\otimes\ket{0}\otimes\ket{2} - \ket{2}\otimes\ket{2}\otimes\ket{0})\nonumber\\
+ \sqrt{\frac{1 -\lambda_4}2}(\ket{0}\otimes\ket{1}\otimes\ket{2} - \ket{0}\otimes\ket{2}\otimes\ket{1}).
\end{eqnarray}
\end{subequations}

Another example will be given in Section~\ref{sec::EM} in the context of estimating entanglement of the constructed subspaces.

The disadvantage of the described approach is that we cannot construct genuinely entangled subspaces of maximal dimension -- the resulting subspace has the same dimension as the initial bipartite one. In general, according to Parthasarathy's result \cite{Parth04}, the maximal dimension of completely entangled subspaces of a $d_1\otimes d_2$ Hilbert space is 
\begin{equation}\label{MaxDim}
d_1 d_2 - (d_1 + d_2) + 1,
\end{equation}
and so this is the upper bound for the dimension of  genuinely entangled $d_1\otimes d'_2\otimes d'_3$ subspaces generated with this approach ($d'_2$, $d'_3$ -- output dimensions of the isometry used for the construction).

\subsection{\label{sec::mxdim}Constructing GES of maximal dimension}

\begin{figure}[t]
\includegraphics[scale=0.35]{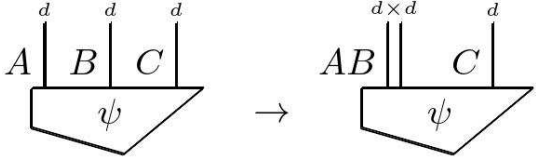}% Here is how to import EPS art
\caption{\label{fig:gather} Gathering subsystems of a tripartite pure state $\ket{\psi}_{ABC}$ into bipartition $AB|C$. The dimensions of the corresponding Hilbert spaces are shown.}
\end{figure}

Our next approach aims at constructing genuinely entangled subspaces of maximal possible dimension.
Given a pure state $\ket{\psi}_{ABC}$ on a tripartite $d\otimes d\otimes d$ Hilbert space, we consider its bipartition $AB|C$ where subsystems $A$ and $B$ are gathered into one subsystem $AB$ described by a Hilbert space of dimension $d^2$~(see Fig.~\ref{fig:gather}).

\begin{figure}[t]
\includegraphics[scale=0.35]{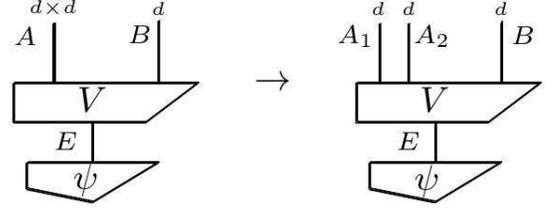}% Here is how to import EPS art
\caption{\label{fig:isogen}Steps of constructing a genuinely entangled subspace: the isometry $V$ maps a Hilbert space $H_E$ to a completely entangled subspace of $H_A\otimes H_B$, then subsystem $A$ is factorized into $A_1$ and $A_2$.}
\end{figure}

According to Ref.~\cite{Parth04}, the maximal dimension of subspaces, which are completely entangled over bipartition $AB|C$, is $d^2\times d - (d^2 +d) +1$. The same being said about the other two bipartitions, $BC|A$ and $AC|B$, this number is the maximal dimension of genuinely entangled $d\otimes d\otimes d$ subspaces.

Our strategy is first to construct a maximal dimensional completely entangled  subspace of a tensor product Hilbert space $H_A\otimes H_B$ with $\mathrm{dim}(H_A) = d^2$ and $\mathrm{dim}(H_B) = d$, then \emph{factorize} the first subsystem $A$ into two $d$-dimensional subsystems $A_1$ and $A_2$, i.~e., perform an  operation somewhat inverse to the one shown on Fig.~\ref{fig:gather}. There are many ways to do such a factorization, and  for our purpose it should be chosen in such a way that the resulting tripartite states are entangled  across bipartitions $A_1|A_2B$ and $A_2|A_1B$ (the entanglement across $A_1A_2|B$ is preserved under such an operation).

According to the above plan, we first construct an isometry $V\colon\,H_E\rightarrow H_A\otimes H_B$, where $H_E$ -- a Hilbert space of dimension $d^2\times d - (d^2 +d) +1$. If this isometry corresponds, via Stinespring's dilation, to a quantum channel with output purity less than one, then its range will be a completely entangled subspace of $H_A\otimes H_B$ with maximal dimension. Next, we apply  a proper factorization of $H_A$ into $H_{A_1}\otimes H_{A_2}$ and obtain a genuinely entangled tripartite subspace of maximal possible dimension, $d^2\times d - (d^2 +d) +1$~(see also Fig.~\ref{fig:isogen}). 

In order to find the desired isometry, we analyze the expression for the output purity of a quantum channel $\mathrm\Phi\colon\,\mathcal B(H_E)\rightarrow\mathcal B(H_A)$ in terms of its Kraus operators. As it was mentioned above, it will suffice to consider the action of $\mathrm\Phi$ only on pure states, and for an arbitrary pure state $\ket{\psi}\in H_E$ we have:
\begin{eqnarray}
\mathrm{Tr}(\mathrm\Phi (\dyad{\psi})^2)&&\nonumber\\  =&&\sum_{i,\,j}\mathrm{Tr}\left(K_i\dyad{\psi}K_i^{\dagger}K_j\dyad{\psi}K_j^{\dagger}\right)\nonumber\\
=&&\sum_{i,\,j}\mathrm{Tr}\left(\dyad{\psi}K_i^{\dagger}K_j\dyad{\psi}K_j^{\dagger}K_i\right)\nonumber\\
=&&\sum_{i,\,j}\abs{\bra{\psi}K_i^{\dagger}K_j\ket{\psi}}^2\nonumber\\
\leqslant&&\sum_{i,\,j}\bra{\psi}K_i^{\dagger}K_i\ket{\psi}\bra{\psi}K_j^{\dagger}K_j\ket{\psi}\nonumber\\
&&=1\label{KrT},
\end{eqnarray}
where the Kraus decomposition~(\ref{Kraus}) was used; the inequality follows from the Cauchy-Schwarz inequality; the last equality follows from eq.~(\ref{TrP}) averaged over $\ket{\psi}$ and squared.

It is known that the Cauchy-Schwarz inequality for two vectors holds strictly, without equality if and only if they are linearly independent. Therefore, to guarantee that the trace in Eq.~(\ref{KrT}) is strictly less than one, we need to search for Kraus operators $\{K_i\}$ satisfying the following conditions:
\begin{enumerate}
    \item The eigenvalues of $K_i^{\dagger}K_i$, for all $i$, are strictly less than one~(otherwise, the trace in Eq.~(\ref{KrT}) will evaluate to $1$ on the corresponding eigenvector)):
    \begin{equation}\label{Kc1}
        \norm{K_i^{\dagger}K_i} < 1.
    \end{equation}
    \item For \emph{any} $\ket{\psi}\in H_E$ there is at least one pair $\left(K_i,\,K_j\right)$ of distinct Kraus operators such that the vectors $K_i\ket{\psi}$ and $K_j\ket{\psi}$ are not proportional to each other.
\end{enumerate}

Having found the proper Kraus operators $\{K_i\}$, we construct the corresponding isometry $V$ via eq.~(\ref{IsoK}). The final step is to do an appropriate factorization of the fist subsystem into two ones, and this procedure significantly depends on the structure of the completely entangled subspace determined by the isometry $V$. 

We illustrate the approach with several examples.

\subsubsection{GES in three-qubit systems}

Now we consider a system of $3$ qubits whose pure states are described by vectors in a $2\otimes 2\otimes 2$ Hilbert space. According to the described above procedure, we gather the first and the second subsystems into one and try to construct a completely entangled $4\otimes 2$ subspace of maximal dimension $4\times 2 - (4 + 2) + 1 = 3$. The corresponding isometry $V\colon\,H_E\rightarrow H_A\otimes H_B$, with $\mathrm{dim}(H_E) = 3$, $\mathrm{dim}(H_A) = 4$, $\mathrm{dim}(H_B) = 2$, can be expressed by eq.~(\ref{IsoK}) in terms of Kraus operators $K_i\colon\,H_E\rightarrow H_A$ of a quantum channel with output purity less than one. 

There are two Kraus operators in this particular case, and, in search for $K_1,\,K_2$, we can consider the simplest situation when each $K_i^{\dagger}K_i$ is diagonal, with some numbers $\lambda^{(i)}_1,\,\lambda^{(i)}_2,\,\lambda^{(i)}_3$ on the main diagonal, such that
\begin{eqnarray}\label{lam22}
    &0 < \lambda^{(i)}_j < 1,\quad  \lambda^{(1)}_j + \lambda^{(2)}_j = 1&,\nonumber\\
    &i=1,\,2;\,j=1,\,2,\,3.&
\end{eqnarray}
Conditions~(\ref{TrP}) and~(\ref{Kc1}) are then satisfied. Such a choice suggests the following structure of $\{K_i\}$ in the form of singular value decomposition:
\begin{equation}\label{K42}
 K_i = W_i   \begin{pmatrix}
    \sqrt{\lambda^{(i)}_1} &          0      & 0\\
          0       &   \sqrt{\lambda^{(i)}_2}  & 0\\
          0       &      0          & \sqrt{\lambda^{(i)}_3}\\
          0       &      0          &   0
    \end{pmatrix},
\end{equation}
where there is some freedom in the choice of  unitary $4\times 4$ matrices $W_i$, $i=1,\,2$.

Trying to keep everything as simple as possible, we can set $W_1 = I$, and choose $W_2$ to be some permutation matrix, for example, the one corresponding to a permutation \begin{equation*}\left(\begin{array}{cc}
     1\, 2\, 3\, 4  \\
     4\, 1\, 2\, 3 
\end{array}\right)
\end{equation*}
Now let us choose an arbitrary vector state $\ket{\phi}\in H_E$ with components $(\phi_1,\,\phi_2,\,\phi_3)$ and, using eq.~(\ref{K42}) and our particular choice of $W_1,\,W_2$,  write out the corresponding vectors $K_1\ket{\phi}$
and $K_2\ket{\phi}$ as columns of a matrix:
\begin{equation}
    \begin{pmatrix}
    \sqrt{\lambda^{(1)}_1}\phi_1 & \sqrt{\lambda^{(2)}_2}\phi_2\\
    \sqrt{\lambda^{(1)}_2}\phi_2 & \sqrt{\lambda^{(2)}_3}\phi_3\\
    \sqrt{\lambda^{(1)}_3}\phi_3 & 0\\
    0 & \sqrt{\lambda^{(2)}_1}\phi_1
    \end{pmatrix}.
\end{equation}
It is easily seen that these vectors are linearly independent for any nonzero $\ket{\phi}$: all $2\times 2$ minors of the matrix evaluate to $0$ simultaneously if and only if \mbox{$\phi_1 = \phi_2 = \phi_3 = 0$}. Therefore, we have shown that operators $K_1,\,K_2$ are the Kraus operators of a channel with output purity less than one. Reconstructing the associated isometry $V\colon\,H_E\rightarrow H_A\otimes H_B$ with the use of eq.~(\ref{IsoK}) and acting with it on an orthonormal basis in $H_E$, we obtain an orthonormal system of vectors spanning a completely entangled subspace of $H_A\otimes H_B$:
\begin{subequations}\label{ces22}
\begin{equation}
    \sqrt{\lambda^{(1)}_1}\ket{0}\otimes\ket{0} + \sqrt{\lambda^{(2)}_1}\ket{3}\otimes\ket{1},
\end{equation}
\begin{equation}\label{2q2}
    \sqrt{\lambda^{(1)}_2}\ket{1}\otimes\ket{0} + \sqrt{\lambda^{(2)}_2}\ket{0}\otimes\ket{1},
\end{equation}
\begin{equation}
    \sqrt{\lambda^{(1)}_3}\ket{2}\otimes\ket{0} + \sqrt{\lambda^{(2)}_3}\ket{1}\otimes\ket{1}.
\end{equation}
\end{subequations}
Next, we try to convert this subspace into a genuinely entangled tripartite one by choosing a proper way of factorization of the first subsystem $A$ of dimension $4$ into subsystems $A_1$ and $A_2$, each with dimension $2$. A naive approach would be to choose the following correspondence between the bases of $H_A$ and $H_{A_1}\otimes H_{A_2}$:
\begin{eqnarray}\label{schq}
\ket{0}\qquad&\rightarrow&\qquad\ket{0}\otimes\ket{0},\nonumber\\
\ket{1}\qquad&\rightarrow&\qquad\ket{0}\otimes\ket{1},\nonumber\\
\ket{2}\qquad&\rightarrow&\qquad\ket{1}\otimes\ket{0},\nonumber\\
\ket{3}\qquad&\rightarrow&\qquad\ket{1}\otimes\ket{1},
\end{eqnarray}
but, obviously, after such a replacement, the spanning vector~(\ref{2q2}) itself turns into a separable one. Any permutation of scheme~(\ref{schq}) doesn't help either.

\begin{figure}[b]
\includegraphics[scale=0.35]{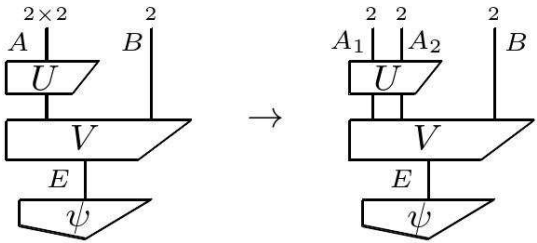}% Here is how to import EPS art
\caption{\label{fig:isounitq} Construction of a genuinely entangled $2\otimes 2\otimes 2$ subspace. A properly chosen unitary $U$ creates entanglement across bipartitions $A_1|A_2B$ and $A_2|A_1B$ while preserving entanglement across $A_1A_2|B$. Line splitting denotes factorization of subsystem $A$ according to scheme~(\ref{schq}). }
\end{figure}

Our second approach is to act on subsystem $A$ with some simple unitary transformation $U_A$ and then factorize $A$ according to scheme~(\ref{schq}). $U_A$  is a local transformation with respect to bipartition $A_1A_2|B$ and, therefore, it preserves entanglement here. On the other hand, $U_A$ can change the situation across bipartitions $A_1|A_2B$ and $A_2|A_1B$ and create entanglement there~(see also Fig.~\ref{fig:isounitq}). We choose this transformation to be
\begin{equation}\label{uniq}
    U_A = \begin{pmatrix}
    1 & 0 & 0 & 0\\
    0 & \frac1{\sqrt{2}} & \frac1{\sqrt{2}} & 0\\
    0 & \frac1{\sqrt{2}} & -\frac1{\sqrt{2}} & 0\\
    0 & 0 & 0 & 1
    \end{pmatrix},
\end{equation}
thus mixing $\ket{1}_A$ and $\ket{2}_A$.

Combining unitary transformation~(\ref{uniq}) and factorization~(\ref{schq}), from the basis~(\ref{ces22}) we obtain an orthonormal system of vectors in $H_{A_1}\otimes H_{A_2}\otimes H_B$:
\begin{subequations}\label{ges22}
\begin{eqnarray}
\ket{\psi_1} &=& \sqrt{\lambda_1}\ket{0}\otimes\ket{0}\otimes\ket{0}\nonumber\\
& &+ \sqrt{1 - \lambda_1}\ket{1}\otimes\ket{1}\otimes\ket{1},\\
\ket{\psi_2} &=& \sqrt{\frac{\lambda_2}{2}}\left(\ket{0}\otimes\ket{1} + \ket{1}\otimes\ket{0}\right)\otimes\ket{0}\nonumber\\ 
& &+ \sqrt{1 - \lambda_2}\ket{0}\otimes\ket{0}\otimes\ket{1},\\
\ket{\psi_3} &=& \sqrt{\frac{\lambda_3}{2}}\left(\ket{0}\otimes\ket{1} - \ket{1}\otimes\ket{0}\right)\otimes\ket{0}\nonumber\\
&+& \sqrt{\frac{1 - \lambda_3}{2}}\left(\ket{0}\otimes\ket{1} + \ket{1}\otimes\ket{0}\right)\otimes\ket{1},
\end{eqnarray}
\end{subequations}
where eq.~(\ref{lam22}) is taken into account and the redundant superscrits of lambdas are now omitted.

We already know that the subspace spanned by vectors~(\ref{ges22}) is completely entangled across bipartition $A_1A_2|B$. The entanglement across other bipartitions, for example, $A_1|A_2B$, can be certified in a standard way --  vectors~(\ref{ges22}) are decomposed accordingly
\begin{equation}
    \ket{\psi_{\mu}} = \sum_{i,\,j,\,k}\,a^{(\mu)}_{i,\,jk}\ket{i}_{A_1}\otimes\ket{jk}_{A_2B},\quad\mu = 1,\,2,\,3,
\end{equation}
and it is analyzed at what complex values $\beta_{\mu}$ the linear combination of matrices
\begin{equation}
    \sum_{\mu=1}^3\,\beta_{\mu}a^{(\mu)}
\end{equation}
has rank less than two, i.~e., all minors of order two evaluate to zero. If there is only trivial solution, $\beta_{\mu}=0\quad\forall\,\mu$, then the subspace is entangled across the cut. This can be done by hand, or, even better, with a computer algebra system supporting the Groebner basis algorithm~\cite{Buch70}.

Using the described procedure, it can be verified that the subspace determined by eq.~(\ref{ges22}) is completely entangled across bipartitions $A_1|A_2B$ and $A_2|A_1B$ and, therefore, is genuinely entangled.

\begin{figure}[b]
\includegraphics[scale=0.35]{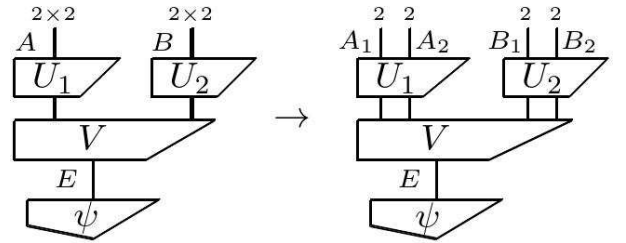}% Here is how to import EPS art
\caption{\label{fig:fq} Construction of a genuinely entangled $2\otimes2\otimes2\otimes2$ subspace. A properly chosen unitaries $U_1$ and $U_2$ create entanglement  across all bipartitions except $A_1A_2|B_1B_2$. Line splittings denote factorizations of subsystems $A$ and $B$ according to scheme~(\ref{schq}). }
\end{figure}

\subsubsection{GES in four-qubit systems}

Pure states of a four-qubit system are described by vectors in a $2\otimes2\otimes2\otimes2$ Hilbert space. Let $A_1$, $A_2$, $B_1$, $B_2$ denote the $4$ subsystems. Joining $3$ subsystems, say $A_2$, $B_1$, $B_2$, into one, we consider the bipartition $A_1|A_2B_1B_2$ and  vector states in a $2\otimes8$ Hilbert space, where the maximal dimension of CES  is equal to $2\times8-(2+8)+1=7$. Bipartitions of another type, such as $A_1A_2|B_1B_2$, yield the maximal dimension of CES equal to $4\times4-(4+4)+1=9$. Therefore, the maximal dimension of GES in a $2\otimes2\otimes2\otimes2$ Hilbert space is upper-bounded by $7$. 

To consruct such GES, we first construct a $7$-dimensional CES in a $4\otimes4$ Hilbert space, then factorize each four-dimensional subsystem into two two-dimensional ones. 

To obtain the CES, we search for the isometry $V\colon\,H_E\rightarrow H_A\otimes H_B$, with the input dimension $\mathrm{dim}(H_E) = 7$ and the output dimensions $\mathrm{dim}(H_A) = 4$, $\mathrm{dim}(H_B) = 4$. In addition, the range of $V$  must be completely entangled.

The subsystems $A$ and $B$ are then factorized into $A_1,\,A_2$ and $B_1,\,B_2$, respectively, and each factorization is done according to scheme~(\ref{schq}). This is still not sufficient for obtaining a GES, and the process must be accompanied with applying some unitary transformations $U_1$ and $U_2$ to the parties $A$ and $B$. $U_1$ and $U_2$ are aimed at creating entanglement across all bipartitions except $A_1A_2|B_1B_2$ where entanglement is guaranteed by the structure of the isometry $V$. The whole procedure is presented on Fig.~\ref{fig:fq}.

The $7$-dimensional CES of a $4\otimes4$ Hilbert space can be obtained by the  approach used in the previous examples. By our construction~(see Appendix~\ref{app:44} for the details), it is spanned by the orthonormal system
\begin{subequations}\label{sfs}
\begin{equation}
    \sqrt{\lambda_1}\ket{0}\otimes\ket{0} + \sqrt{1-\lambda_1}\ket{2}\otimes\ket{1}, 
\end{equation}
\begin{equation}
    \sqrt{\lambda_2}\ket{1}\otimes\ket{2} + \sqrt{1-\lambda_2}\ket{3}\otimes\ket{3}, 
\end{equation}
\begin{equation}
    \sqrt{\lambda_3}\ket{0}\otimes\ket{1} + \sqrt{1-\lambda_3}\ket{3}\otimes\ket{2}, 
\end{equation}
\begin{equation}
    \sqrt{\lambda_4}\ket{3}\otimes\ket{0} + \sqrt{1-\lambda_4}\ket{2}\otimes\ket{2}, 
\end{equation}
\begin{equation}
    \sqrt{\lambda_5}\ket{1}\otimes\ket{1} + \sqrt{1-\lambda_5}\ket{0}\otimes\ket{3}, 
\end{equation}
\begin{equation}
    \sqrt{\lambda_6}\ket{1}\otimes\ket{0} + \sqrt{1-\lambda_6}\ket{2}\otimes\ket{3}, 
\end{equation}
\begin{equation}
    \sqrt{\lambda_7}\ket{2}\otimes\ket{0} + \sqrt{1-\lambda_7}\ket{3}\otimes\ket{1}, 
\end{equation}
\end{subequations}
where $0<\lambda_i<1,\:i=1,\ldots,7$.

The factorization by scheme~(\ref{schq}) of each of the two subsystems inevitably creates separable vectors from those in eq.~(\ref{sfs}), and so we need some entangling unitaries $U_1$ and $U_2$ to apply to $A$ and $B$, respectively. There are $7$ possible bipartitions of a four-partite system, and, choosing $U_1$ and $U_2$, we need to check entanglement in $6$ of them. Entanglement across bipartition $A_1A_2|B_1B_2$ is provided by the structure of the isometry $V$. Of course, such verification can be most effectively done with the use of computer algebra and the Groebner basis algorithm.

Due to the computational complexity of the problem, we were able to calculate only with concrete values for $\lambda_i$ from eqs.~(\ref{sfs}). At first we set $\lambda_i = 1/2$, $i=1,\ldots,7$. In addition, unlike the example with $3$ qubits, we couldn't come up with simple unitaries $U_1$ and $U_2$ mixing only $2$-$3$ basis states. From Ref.~\cite{Tremain2011} we took several standard unitaries, $3$- and $4$-dimensional, and on the basis of them constructed two $4$-dimensional unitaries for mixing basis states:
\begin{equation}
    Q=\frac12\begin{pmatrix}
    -1 & 1  &  1 &  1\\
     1 & -1 &  1 &  1\\
     1 &  1 & -1 &  1\\
     1 &  1 &  1 & -1
    \end{pmatrix},
\end{equation}
\begin{equation}
    T = \begin{pmatrix}
    -1/3 & 2/3 & 0 & 2/3\\
    2/3 & -1/3 & 0 & 2/3\\
    0 & 0 & 1 & 0\\
    2/3 & 2/3 & 0 & -1/3
    \end{pmatrix}.
\end{equation}
Setting then
\begin{equation}
    U_1 = QT,\qquad U_2 = T
\end{equation}
did the job.

Combining these transformations with factorization~(\ref{schq}), we obtain two new schemes of factorization. For subsystem $A$ it is defined by
\begin{eqnarray}\label{subA}
\ket{0}_A\quad&\rightarrow&\quad\frac16(5\ket{0}\otimes\ket{0} - \ket{0}\otimes\ket{1}\nonumber\\
&+&3\ket{1}\otimes\ket{0} - \ket{1}\otimes\ket{1}),\nonumber\\
\ket{1}_A\quad&\rightarrow&\quad\frac16(-\ket{0}\otimes\ket{0} + 5\ket{0}\otimes\ket{1}\nonumber\\
&+&3\ket{1}\otimes\ket{0} - \ket{1}\otimes\ket{1}),\nonumber\\
\ket{2}_A\quad&\rightarrow&\quad\frac12(\ket{0}\otimes\ket{0} + \ket{0}\otimes\ket{1}\nonumber\\
&-&\ket{1}\otimes\ket{0} + \ket{1}\otimes\ket{1}),\nonumber\\
\ket{3}_A\quad&\rightarrow&\quad\frac16(-\ket{0}\otimes\ket{0} - \ket{0}\otimes\ket{1}\nonumber\\
&+&3\ket{1}\otimes\ket{0} + 5\ket{1}\otimes\ket{1}),
\end{eqnarray}
for subsystem $B$ it is given by
\begin{eqnarray}\label{subB}
\ket{0}_B\quad&\rightarrow&\quad\frac13(-\ket{0}\otimes\ket{0} + 2\ket{0}\otimes\ket{1}\nonumber\\
&&+ 2\ket{1}\otimes\ket{1}),\nonumber\\
\ket{1}_B\quad&\rightarrow&\quad\frac13(2\ket{0}\otimes\ket{0} - \ket{0}\otimes\ket{1}\nonumber\\
&&+ 2\ket{1}\otimes\ket{1}),\nonumber\\
\ket{2}_B\quad&\rightarrow&\quad\ket{1}\otimes\ket{0},\nonumber\\
\ket{3}_B\quad&\rightarrow&\quad\frac13(2\ket{0}\otimes\ket{0} + 2\ket{0}\otimes\ket{1}\nonumber\\
&&- \ket{1}\otimes\ket{1}).
\end{eqnarray}

Now, to obtain a four-qubit GES, one should apply substitutions~(\ref{subA}) and (\ref{subB}) in eqs.~(\ref{sfs}) for the basis states of the first and the second subsystems, respectively, and set all $\lambda_i$ to $1/2$. 

Actually, we've done calculations for several different sets of values of $\lambda_i$ and each time obtained a GES. 
We can conjecture that eqs.~(\ref{sfs}) together with eqs.~(\ref{subA}) and (\ref{subB}) determine a GES for \emph{any} set of values of $\lambda_i$.

\subsubsection{\label{subsub:333}GES in three-qutrit systems}

Now we construct an example of GES of maximal dimension in a $3\otimes 3\otimes 3$ Hilbert space. The bound on the  dimension in this case is the same as the bound on the dimension  of a $9\otimes 3$ CES: $9\times3 - (9 + 3) + 1 = 16$. Consequently, we need to construct an isometry $V\colon\,H_E\rightarrow H_A\otimes H_B$, with $\mathrm{dim}(H_E) = 16$, $\mathrm{dim}(H_A) = 9$, $\mathrm{dim}(H_B) = 3$, and with a completely entangled range. 

Our first step is to construct the Kraus operators $K_i\colon\,H_E\rightarrow H_A,\,i=1,\,2,\,3$, which in this case are represented by  matrices with dimensions $9\times16$. As in the previous examples, we first  consider matrices corresponding to $K_i^{\dagger}K_i$, which should satisfy eqs.~(\ref{TrP}) and~(\ref{Kc1}). Any attempt to make them all be purely diagonal immediately fails:  we can write out the diagonals of $K_1^{\dagger}K_1$, $K_2^{\dagger}K_2$, $K_3^{\dagger}K_3$ as columns of a matrix and, to satisfy eqs.~(\ref{TrP}) and~(\ref{Kc1}), in each row of this matrix there must be at least two nonzero values adding up to $1$. Therefore, the total number of these values in all rows must be  not less than $16\times 2=32$. On the other hand, being the eigenvalues of $K_i^{\dagger}K_i$, these numbers are the squares of singular values of $K_i$. There can be at most $9$ nonzero singular values for each $K_i$ because it is represented by a $9\times 16$ matrix. In total, for all three $K_i$ there can be at most $3\times 9 = 27$ nonzero singular values, so we have a contradiction which shows that $K_i^{\dagger}K_i$ cannot be purely diagonal. 

To come around the described problem, we introduce a block-diagonal form for $K_i^{\dagger}K_i$. For each matrix representing $K_i^{\dagger}K_i$ we consider the area consisting of three diagonals: the main diagonal and the two adjacent ones below and above. This area is divided into two parts down the diagonals: the fist part entirely consists of $2\times 2$ blocks and the second one is purely diagonal -- it has nonzero values only on the main diagonal. The $2\times 2$ blocks placed on the same positions in $K_1^{\dagger}K_1$, $K_2^{\dagger}K_2$, $K_3^{\dagger}K_3$ should add up to a $2\times 2$ identity matrix to satisfy eq.~(\ref{TrP}). To satisfy also eq.~(\ref{Kc1}), they should have eigenvalues strictly less than $1$. As we've shown above, for each $K_i^{\dagger}K_i$ we only have $9$ eigenvalues at our disposal to cover the diagonals, so we need to spare them. This can be achieved by choosing all $2\times 2$ blocks to have only $1$ nonzero eigenvalue. In this case each such block will cover two positions on the diagonal, but it will take only one eigenvalue. Having placed $5$ such blocks down the diagonals in each $K_i^{\dagger}K_i$, we will have $9 - 5 = 4$ eigenvalues left at our disposal. We can then put these eigenvalues down the main diagonal, thus covering the remaining positions. If we put the described three-diagonal areas of $K_1^{\dagger}K_1$, $K_2^{\dagger}K_2$, $K_3^{\dagger}K_3$ in front of each other in the form of columns, the picture will look like this:
\begin{equation}\label{Kstruct}
    \begin{matrix}
    K_1^{\dagger}K_1\colon & & & &   K_2^{\dagger}K_2\colon  & & & & K_3^{\dagger}K_3\colon\\
    \\
    P_1^{(1)}& & & & P_1^{(2)} & & & & P_1^{(3)}\\
    P_2^{(1)}& & & & P_2^{(2)} & & & & P_2^{(3)}\\
    P_3^{(1)}& & & & P_3^{(2)} & & & & P_3^{(3)}\\
    P_4^{(1)}& & & & P_4^{(2)} & & & & P_4^{(3)}\\
    P_5^{(1)}& & & & P_5^{(2)} & & & & P_5^{(3)}\\
    \lambda_1^{(1)} & & & & 0 & & & & \lambda_1^{(3)}\\
    0 & & & & \lambda_1^{(2)} & & & & \lambda_2^{(3)}\\
    \lambda_2^{(1)} & & & & \lambda_2^{(2)} & & & & 0\\
    \lambda_3^{(1)} & & & & 0 & & & & \lambda_3^{(3)}\\
     0 & & & & \lambda_3^{(2)} & & & & \lambda_4^{(3)}\\
     \lambda_4^{(1)} & & & & \lambda_4^{(2)} & & & & 0
    
    \end{matrix},
\end{equation}
where lambda's in each row  are positive and add up to $1$, and $P_i^{(j)}$ -- $2\times 2$ matrices having only $1$ nonzero~(positive) eigenvalue and adding up to a $2\times2$ identity matrix:
\begin{equation}\label{povm}
    P_i^{(1)} + P_i^{(2)} + P_i^{(3)} = I,\quad i=1,\dots,5.
\end{equation} 

Any other arrangement of lambda's is acceptable as long as there are at least two of them adding up to $1$ in each row. Here we've placed them in some checkerboard style.

The number of $2\times 2$ blocks used in each $K_i^{\dagger}K_i$ is optimal: if we had used $4$ blocks instead, then we would have had $9-4=5$ free lambda's to cover $16 - 4\times2 = 8$ positions on the diagonal, which is impossible even with the use of such a sparse checkerboard disposition.

The operators satisfying eq.~(\ref{povm}) for each fixed $i$ constitute a POVM in a two-dimensional Hilbert space. One particular choice of $P^{(k)}$~\cite{Preskill}  can be defined by 
\begin{eqnarray}\label{PrP}
P^{(1)} = A\dyad{u},\: P^{(2)}=A\dyad{v}&&\nonumber\\
P^{(3)} = I - A(\dyad{u} + \dyad{v}),&&
\end{eqnarray}
where $A$ -- a constant and $\ket{u}$, $\ket{v}$ -- vectors parameterized by
\begin{eqnarray}
    &\ket{u}=\left(\begin{array}{c}
         \cos{\alpha}  \\
         \sin{\alpha} 
    \end{array}
    \right),\quad\ket{v}=\left(\begin{array}{c}
         \sin{\alpha}  \\
         \cos{\alpha} 
    \end{array}
    \right)&,\nonumber\\
    &0 < \alpha < \pi/4.&
\end{eqnarray}

It can be easily verified that when 
\begin{equation}\label{Ac}
    A = \frac1{1+\sin{2\alpha}},
\end{equation}
the operator $P^{(3)}$ has only one nonzero eigenvalue, and thus the identity operator $I$ is decomposed into the sum of three rank one operators $P^{(k)}$.

By construction~(\ref{Kstruct}), there are five triples of operators $\left(P_i^{(1)},\,P_i^{(2)},\,P_i^{(3)}\right)$, $i = 1,\ldots,5$, and each is defined by its own parameter $\alpha_i$ according to eqs.~(\ref{PrP})~-~(\ref{Ac}). From now on we will denote such triples as $\left(P_{\alpha_i}^{(1)},\,P_{\alpha_i}^{(2)},\,P_{\alpha_i}^{(3)}\right)$, $i = 1,\ldots,5$, thus stressing out their dependence on $\alpha_i$.

Having found proper $K_i^{\dagger}K_i$ with the use of eqs.~(\ref{Kstruct})~-~(\ref{Ac}), we can construct operators $K_i$ themselves taking advantage of the singular value decomposition:
\begin{equation}\label{svd333}
    K_i = W_i\,\Sigma_i\,V_i,\quad i=1,\,2,\,3,
\end{equation}
where $W_i$ is a $9\times9$ unitary; $\Sigma_i$ is a $9\times16$ matrix consisting of $\sqrt{\lambda^{(i)}_j}$ and square roots of the eigenvalues of the operators $P^{(i)}_j$; $V_i$~is~a $16\times 16$ unitary which depends on $\alpha_j$, $j=1,\ldots,5$~(see Appendix~\ref{app:333} for details).

There is  freedom in choosing the unitaries $W_i$, which we set to  some permutation matrices. 

When we set particular values for $\lambda^{(i)}_j$ and $\alpha_j$, the procedure of construction of a $9\otimes3$ CES reduces to the search of appropriate permutations $W_i$. According to the general approach, we choose an arbitrary vector state $\ket{\phi}\in H_E$ with components $\phi_j$, $j=1,\ldots,16$, and write the corresponding vectors $K_i\ket{\phi}$, $i=1,\,2,\,3$,
as columns of a matrix. If there is only trivial solution $\{\phi_j\}$, for which all $2\times2$ minors of the matrix evaluate to zero, then at least two of these vectors are linearly independent, and we have a CES. We use a computer algebra system and the Groebner basis algorithm for such an analysis. When Groebner basis reduces to $\{1\}$, there is only trivial solution. If it is not the case, we look at the basis, identify  the minors which  cause  problem and  adjust the permutations $W_i$ accordingly~(usually it takes one or two elementary transpositions). After that we compute the Groebner basis again and, if necessary, repeat the procedure until we obtain a CES.

We were not able to calculate in abstract variables here due to high dimensionality of the problem and the complexity of the polynomial system for minors, so we  made a particular choice of $\lambda^{(i)}_j$ and $\alpha_j$ for scheme~(\ref{Kstruct})~(see Appendix~\ref{app:333}, eq.~(\ref{InSet})). For such initial setting we came up with the proper permutations $W_i$~(see Appendix~\ref{app:333}, eq.~(\ref{WP})) in several iterations of the described above procedure. Calculating $K_i$ from eq.~(\ref{svd333}) and constructing the corresponding isometry with the use of eq.~(\ref{IsoK}), we obtained an orthonormal system of vectors spanning a $9\otimes3$ CES~(see Appendix~\ref{app:333}, eq.~(\ref{Sp333})). Next, we factorized the $9$-dimensional subsystem into two $3$-dimensional ones according to some particular scheme~(see Appendix~\ref{app:333}, eq.~(\ref{sch3})). Finally, after the analysis of the two new bipartitions, we introduced a simple unitary transformation acting on the two subsystems and creating entanglement~(see Appendix~\ref{app:333}, eq.~(\ref{U3})). In this way we obtained a $3\otimes3\otimes3$ GES.

\section{\label{sec::EM}Entanglement of GES and the associated bounds on entanglement measures}

\subsection{Controlling genuine entanglement of constructed subspaces}

Let us return to the first approach to constructing GES described in Subsection~\ref{sub::constch}. Such subspaces are obtained from CES of a bipartite Hilbert space by the action of some isometry on one party. The entanglement of GES can be estimated if the characteristics of both the initial CES and the channel corresponding to the isometry are known. Across one bipartition it is determined by the entanglement of the initial CES~(see eq.~(\ref{CESbi}) for bipartition $A|CD$), across the other two -- by the output norm of the quantum channel~(see eq.~(\ref{chbi}) for bipartition $C|AD$). The maximal output norm of the channel will be an upper bound on the norms of the reduced density operators of the states in GES across these two bipartitions because the supremum in eq.~(\ref{OutNorm}) is achieved on pure states.

As an illustration, we take the initial CES to be the antisymmetric subspace $W_-$ of a $3\otimes3$ Hilbert space, spanned by the three vectors
\begin{eqnarray}\label{asym}
    &\ket{\psi}_{ij} =  (\ket{i}\otimes\ket{j} - \ket{j}\otimes\ket{i})/\sqrt2&,\nonumber\\
    &i,\,j = 0,\ldots,2,\quad i<j.&
\end{eqnarray}
It is known~\cite{Vid02},\cite{KVAnt20} that the maximal first Schmidt coefficient  over all states in this subspace is given by
\begin{equation}\label{supsch}
    \sup_{\psi_{AB}\in W_-}\norm{\mathrm{Tr}_B\dyad{\psi_{AB}}}^{1/2} = 1/\sqrt2.
\end{equation}

Consider the Holevo-Werner channel~\cite{HolWern02} acting on $d\times d$ density matrices as
\begin{eqnarray}\label{HW}
\mathrm{\Phi}(\rho) = \frac1{d-1}\left(I - \rho^T\right)\nonumber\\
= \sum_{i<j}K_{ij}\rho K_{ij}^{\dagger},
\end{eqnarray}
where
\begin{eqnarray}\label{KrHW}
    K_{ij} =\frac1{\sqrt{d-1}}\left(|i\rangle\langle j| - |j\rangle\langle i|\right),\nonumber\\
    i,\,j = 0,\ldots,d-1,\quad i<j,
\end{eqnarray}
and $\rho^T$ denotes the matrix transpose with respect to the computational basis.

The maximal output norm of the Holevo-Werner channel can be easily obtained from the first representation in eq.~(\ref{HW}) and is given by
\begin{equation}\label{HWn}
    \nu_p(\mathrm\Phi) = (d-1)^{-(1-1/p)}
\end{equation}

In our case, $d=3$, the channel can be represented by the $3$ Kraus operators given by eq.~(\ref{KrHW}). The associated isometry $V$, obtained from  eq.~(\ref{IsoK}), acts on the computational basis as
\begin{eqnarray}
V\ket{0}&=& \frac1{\sqrt2}\left[-\ket1\otimes\ket0 - \ket2\otimes\ket1\right],\nonumber\\
V\ket{1}&=& \frac1{\sqrt2}\left[\ket0\otimes\ket0 - \ket2\otimes\ket2\right],\nonumber\\
V\ket{2}&=& \frac1{\sqrt2}\left[\ket0\otimes\ket1 + \ket1\otimes\ket2\right].
\end{eqnarray}
Acting with this isometry on the second qutrit of the vectors in eq.~(\ref{asym}), we obtain an orthonormal system spanning a GES of a $3\otimes3\otimes3$ Hilbert space:
\begin{subequations}\label{HWges}
\begin{eqnarray}
\ket{\phi_1} = \frac12(\ket{0}\otimes\ket{0}\otimes\ket{0} - \ket{0}\otimes\ket{2}\otimes\ket{2}\nonumber\\
+ \ket{1}\otimes\ket{1}\otimes\ket{0} + \ket{1}\otimes\ket{2}\otimes\ket{1}),
\end{eqnarray}
\begin{eqnarray}
\ket{\phi_2} = \frac12(\ket{0}\otimes\ket{0}\otimes\ket{1} + \ket{0}\otimes\ket{1}\otimes\ket{2}\nonumber\\
+ \ket{2}\otimes\ket{1}\otimes\ket{0} + \ket{2}\otimes\ket{2}\otimes\ket{1}),
\end{eqnarray}
\begin{eqnarray}
\ket{\phi_3} = \frac12(\ket{1}\otimes\ket{0}\otimes\ket{1} + \ket{1}\otimes\ket{1}\otimes\ket{2}\nonumber\\
- \ket{2}\otimes\ket{0}\otimes\ket{0} + \ket{2}\otimes\ket{2}\otimes\ket{2}).
\end{eqnarray}
\end{subequations}

Let $S$ denote this subspace. Consider the geometric measure $G_{GME}$ of genuine entanglement of $S$. The three qutrits are referred to in the same way as on Figure~\ref{fig:iso23}: $A$, $C$ and $D$. The maximal first Schmidt coefficient over all states in GES across bipartition $A|CD$ is determined by that of the initial antisymmetric bipartite subspace  given by eq.~(\ref{supsch}). The maximal first Schmidt coefficient across bipartitions $C|AD$ and $D|AC$ will be upper bounded by the output norm $\nu_{\infty}(\mathrm\Phi)^{1/2}$ of the Holevo-Werner channel, which, by eq.~(\ref{HWn}), also equals to $1/\sqrt2$. Combining the results across all bipartitions, with the use of eqs.~(\ref{geom}), (\ref{geomGME}) we obtain the lower bound on the geometric measure of $S$:
\begin{equation}\label{Gb}
    G_{GME}(S)\geqslant 1/2.
\end{equation}

\subsection{Lower bounds on GME concurrence and negativity of mixed states}

Obtained in Ref.~\cite{KVAnt20}  lower bounds on the concurrence and the convex-roof extended negativity of an arbitrary bipartite mixed state $\rho$ are determined by its overlap with some completely entangled subspace $W$ of a $d\times d$ Hilbert space. For the concurrence the bound reads
\begin{equation}
    C(\rho)\,\geqslant\,\max\left(\sqrt{\frac{2}{d(d-1)}}\,\frac{\Tr{\rho\,\Pi_W} - \lsup}{\lsup},\,0\right),
\end{equation}
where $\Pi_W$ -- an orthogonal projector onto $W$, $\lsup$ -- the supremum of the largest Schmidt coefficient squared taken over all vector states in the subspace $W$.
The bound on the negativity is given by
\begin{equation}
    N^{\mathrm{CREN}}(\rho)\,\geqslant\,\max\left(\frac{\Tr{\rho\,\Pi_W} - \lsup}{2\lsup},\,0\right).
\end{equation}

The underlying separability criterion was derived earlier in Ref.~\cite{Sar08} with the use of the theory of entanglement witnesses.~(see also Ref.~\cite{DemAugWit18} for the generalization to the GME case).

With the use of eqs.~(\ref{geom})-(\ref{cfGME}) these inequalities can be straightforwardly extended to the GME case:
\begin{eqnarray}
    C_{GME}(\rho)\,&&\geqslant\,\sqrt{\frac{2}{d(d-1)}}\frac1{1 - G_{GME}(W)}\nonumber\\
    \times&&\max(\Tr{\rho\,\Pi_W} + G_{GME}(W) - 1,\,0),\label{Cgme}\\
    N_{GME}(\rho)\,&&\geqslant\,\frac1{2(1 - G_{GME}(W))}\nonumber\\
    \times&&\max(\Tr{\rho\,\Pi_W} + G_{GME}(W) - 1,\,0)\label{Ngme},
\end{eqnarray}
where now $W$ is a genuinely entangled subspace. Therefore, the two GME measures can be estimated for a mixed state $\rho$ if the value $G_{GME}(W)$ or some good lower bound on it can be determined.

As an example, with the GES $S$ from eq.~(\ref{HWges}), for a  mixed state $\rho$  the lower bound on the negativity is given by
\begin{equation}
    N_{GME}(\rho)\,\geqslant\,\max(\Tr{\rho\,\Pi_S} - 1/2,\,0),
\end{equation}
where eqs.~(\ref{Gb}) and (\ref{Ngme}) were used.

Eqs.~(\ref{Cgme}) and (\ref{Ngme})~(or the underlying separability criterion) are also convenient for estimating the robustness of genuine multipartite entanglement under  mixing with noise. Consider a state $\rho$ with support in some GES $W$ being mixed with noise described by a density operator $N$. After mixing the state is given by the convex sum
\begin{equation}\label{NoisyS}
    (1-p)\rho + p N,\quad p\in[0;1].
\end{equation}
By eqs.~(\ref{Cgme}) and (\ref{Ngme}), this state will still be genuinely entangled if the lower bound on the entanglement measures is positive, i.~e.,
\begin{equation}
    \mathrm{Tr}\{(1-p)\rho\Pi_W + p N\Pi_W\} + G_{GME}(W) > 1.
\end{equation}
Taking into account that $\mathrm{Tr}\{\rho\Pi_W\}=1$~($\rho$ has support in $W$), we obtain the upper bound
\begin{equation}
    p < \frac{G_{GME}(W)}{1 - \mathrm{Tr}\{N\Pi_W\}}
\end{equation}
saying that under such proportions of the noise $N$ the state in eq.~(\ref{NoisyS}) is guaranteed to be genuinely entangled.

Let $H$ denote the Hilbert space of the described states. If only the white noise is considered, 
\begin{equation}
N = \frac1{\mathrm{dim}(H)}I,
\end{equation}
then the bound on $p$ reads
\begin{equation}\label{wnrob}
    p < \frac{G_{GME}(W)}{1 - \mathrm{dim}(W)/\mathrm{dim}(H)}.
\end{equation}
As an example, for the subspace $S$ given by eq.~(\ref{HWges}) the robustness of entanglement under the white noise  can be estimated with the use of eq.~(\ref{Gb}). It is given by $p < 9/16=0.5625$.

Following Ref~\cite{KVAnt20}, it is also interesting to analyze the case when the expression for $N$ is unknown  and only the spectrum of the noise  is given~(\emph{robustness from spectrum}). The immediate generalization of eq.~(55) in Ref.~\cite{KVAnt20} to the GME case is given by
\begin{equation}\label{SpN}
    p < G_{GME}(W)/\norm{N}_{\scriptstyle{(\mathrm{codim}(W))}},
\end{equation}
where $\mathrm{codim}(W) = \mathrm{dim}(H)-\mathrm{dim}(W)$ and $\norm{N}_{(k)}$ is the sum of $k$ largest eigenvalues of $N$~(\emph{the $k$-th Ky Fan norm}~\cite{Bhatia97, HornJohn13}).

Eq.~(\ref{SpN}) may be useful in analysis of quantum states under the influence of random noise where some results on eigenvalues of random matrices can be applied.

\section{\label{sec::conc}Discussions}

Several methods of constructing GESs have been presented. In the first one tripartite GESs are obtained from bipartite CESs by applying an isometry corresponding to a quantum channel with the maximal output norm strictly less than $1$. On the one hand, the dimensions of the obtained subspaces are not high and bounded by those of the initial CESs. On the other hand, the method is pretty simple and allows us to control the  measures of entanglement of the constructed subspaces on condition that we have control over the CESs and the output characteristics of the channel being used. Highly entangled GESs can be generated in this manner. Quite possibly, other characteristics, such as distillability of entanglement across every bipartition, can be controlled in a similar way. In principle, the whole procedure can be applied to obtain $n+1$-partite GESs from $n$-partite ones. It is also interesting to ask whether this method can be realized in experimental setting: the isometry, for example, can be extended in some way to a unitary which may correspond to some physical operation. These questions will be addressed in the forthcoming work.

The second method is based on the direct construction of certain quantum channels with the subsequent factorization of some compound systems into smaller ones. The method can generate GESs of maximal possible dimension but it gets computationally hard with the increase of the number of parties and the dimensionalities of their systems. We were able to come up with families of maximal GESs for three-qubit systems and, to some extent, for four-qubit and three-qutrit systems where we can only conjecture that we are dealing with  GESs in a wide range of the values of the parameters~(we checked that they are GESs for various sets of the parameters). 

One may benefit from these parameterizations in the analysis of various properties of entangled states and subspaces. As an example, starting from a maximal GES for a system of $3$ qubits given by eqs.~(\ref{ges22}), one can easily find another maximal GES orthogonal to the first one:
\begin{subequations}
\begin{eqnarray}
\ket{\psi_1} &=& \sqrt{1 - \lambda_1}\ket{0}\otimes\ket{0}\otimes\ket{0}\nonumber\\
& &- \sqrt{\lambda_1}\ket{1}\otimes\ket{1}\otimes\ket{1},\nonumber\\
\ket{\psi_2} &=& \sqrt{\frac{1-\lambda_2}{2}}\left(\ket{0}\otimes\ket{1} + \ket{1}\otimes\ket{0}\right)\otimes\ket{0}\nonumber\\ 
& &- \sqrt{\lambda_2}\ket{0}\otimes\ket{0}\otimes\ket{1},\nonumber\\
\ket{\psi_3} &=& \sqrt{\frac{1-\lambda_3}{2}}\left(\ket{0}\otimes\ket{1} - \ket{1}\otimes\ket{0}\right)\otimes\ket{0}\nonumber\\
&-& \sqrt{\frac{\lambda_3}{2}}\left(\ket{0}\otimes\ket{1} + \ket{1}\otimes\ket{0}\right)\otimes\ket{1}.\nonumber
\end{eqnarray}
\end{subequations}
These two GESs span in total $6$ dimensions out of 8 in a tripartite Hilbert space~(but note that their direct sum can't be genuinely entangled). Their orthogonal complement, a $2$-dimensional subspace, contains separable vectors, for example, $\ket{1}\otimes\ket{1}\otimes\ket{0}$.  Given an arbitrary mixed state of $3$ qubits, one can optimize over the parameters $\lambda_i$ and find the GES of one of the two types having maximal overlap with the state. If the state is genuinely entangled, it can be expected that the overlap will be significant. With the geometric measure of the optimal GES obtained, for example, numerically, one could then use eqs.~(\ref{Cgme}), (\ref{Ngme}) to detect genuine entanglement of the state or to get estimates for its entanglement measures. The same can be said in relation to the four-qubit subspaces determined by eqs.~(\ref{sfs}), (\ref{subA}), (\ref{subB}), in which case the two orthogonal GESs will span together $14$ dimensions out of $16$ possible. The numerical analysis of the effectiveness of such an approach  will be conducted elsewhere.

It seems that there is some trade-off between the geometric measure of entanglement of a GES and its dimension. It would be interesting to find examples of GESs optimal with respect to the robustness of entanglement of their states under mixing with the white noise, an estimate for which is given in eq.~(\ref{wnrob}).

\begin{acknowledgments}
The author thanks Otfried G{\"u}hne for many fruitful discussions and valuable comments.
This work was supported by Lomonosov Moscow State University.
\end{acknowledgments}

\appendix

\section{\label{app:33}Construction of a family of $3\otimes3$ CESs of maximal dimension}

The maximal dimension of $3\otimes3$ CES, by eq.~(\ref{MaxDim}), is equal to $4$. According to the theory given in Section~\ref{sec::prel}, we can associate each such CES with some isometry $V\colon\,H_E\rightarrow H_A\otimes H_B$, where $\mathrm{dim}(H_E) = 4$, $\mathrm{dim}(H_A) = 3$, $\mathrm{dim}(H_B) = 3$.

We search for the three Kraus operators $K_i$ of the corresponding channel related to $V$ by eq.~(\ref{IsoK}). To make the subspace be completely entangled, we must satisfy the conditions given in Section~\ref{sec::const} on page~\pageref{Kc1}. At first we consider the operators $K^{\dagger}_i K_i$ which in this case can be constructed in the simplest, diagonal form:
\begin{eqnarray}\label{Kces3}
&K^{\dagger}_1 K_1 = \begin{pmatrix}
\lambda^{(1)}_1 & 0 & 0 & 0\\
0 & \lambda^{(1)}_2 & 0 & 0\\
0 & 0 & \lambda^{(1)}_3 & 0\\
0 & 0 & 0 & 0
\end{pmatrix}&,\nonumber\\
&K^{\dagger}_2 K_2 = \begin{pmatrix}
\lambda^{(2)}_1 & 0 & 0 & 0\\
0 & \lambda^{(2)}_2 & 0 & 0\\
0 & 0 & 0 & 0\\
0 & 0 & 0 & \lambda^{(2)}_4
\end{pmatrix}&,\nonumber\\
&K^{\dagger}_3 K_3 = \begin{pmatrix}
\lambda^{(3)}_1 & 0 & 0 & 0\\
0 & 0 & 0 & 0\\
0 & 0 & \lambda^{(3)}_3 & 0\\
0 & 0 & 0 & \lambda^{(3)}_4
\end{pmatrix}&,
\end{eqnarray}
where
\begin{eqnarray}\label{lam3}
    &0 < \lambda^{(i)}_j < 1,\quad  \lambda^{(1)}_j + \lambda^{(2)}_j + \lambda^{(3)}_j = 1&,\nonumber\\
    &i=1,\ldots,3;\,j=1,\ldots,4&
\end{eqnarray}
to satisfy the conditions of eqs.~(\ref{TrP}) and (\ref{Kc1}).

There can be at most $3$ nonzero values on the main diagonal of each $K^{\dagger}_i K_i$ because each $K_i$ itself is represented by a $3\times4$ matrix that can have at most $3$ nonzero singular values.
In addition, the analysis can be simplified by setting $\lambda^{(2)}_1 = 0$  without violation of the conditions in eq.~(\ref{lam3}).

Eq.~(\ref{Kces3}) suggests that $K_i$ can be written in the form of the singular value decomposition:
\begin{eqnarray}\label{Krces}
&K_1 = W_1\,\begin{pmatrix}
\sqrt{\lambda^{(1)}_1} & 0 & 0 & 0\\
0 & \sqrt{\lambda^{(1)}_2} & 0 & 0\\
0 & 0 & \sqrt{\lambda^{(1)}_3} & 0
\end{pmatrix}&,\nonumber\\
&K_2 = W_2\,\begin{pmatrix}
\sqrt{\lambda^{(2)}_1} & 0 & 0 & 0\\
0 & \sqrt{\lambda^{(2)}_2} & 0 & 0\\
0 & 0 & 0 & \sqrt{\lambda^{(2)}_4}
\end{pmatrix}&,\nonumber\\
&K_3 = W_3\,\begin{pmatrix}
\sqrt{\lambda^{(3)}_1} & 0 & 0 & 0\\
0 & 0 & 0 & \sqrt{\lambda^{(3)}_4}\\
0 & 0 & \sqrt{\lambda^{(3)}_3} & 0
\end{pmatrix},&
\end{eqnarray}
where $W_1$, $W_2$ and $W_3$ -- $3\times3$ unitary matrices which we can choose appropriately to satisfy  our conditions on the Kraus operators. We set $W_1=I$ and choose $W_2$ and $W_3$ to be some permutation matrices:
\begin{equation}
    W_2\colon\,\left(\begin{array}{cc}
     1\, 2\\
     2\, 1 
\end{array}\right),\quad W_3\colon\,
\left(\begin{array}{cc}
     1\, 2\, 3\\
     3\, 1\, 2
\end{array}\right).
\end{equation}

Now we choose an arbitrary vector state $\ket{\phi}\in H_E$ with components $(\phi_1,\,\phi_2,\,\phi_3,\,\phi_4)$ and, using  our particular choice of $W_1,\,W_2$, $W_3$,  write out the corresponding vectors $K_1\ket{\phi}$, $K_2\ket{\phi}$, $K_3\ket{\phi}$ as columns of a matrix:
\begin{equation}
    \begin{pmatrix}
    \sqrt{\lambda^{(1)}_1}\phi_1 & \sqrt{\lambda^{(2)}_2}\phi_2 & \sqrt{\lambda^{(3)}_4}\phi_4\\
    \sqrt{\lambda^{(1)}_2}\phi_2 & 0 & \sqrt{\lambda^{(3)}_3}\phi_3\\
    \sqrt{\lambda^{(1)}_3}\phi_3 & \sqrt{\lambda^{(2)}_4}\phi_4 & \sqrt{\lambda^{(3)}_1}\phi_1
    \end{pmatrix},
\end{equation}
where $\lambda^{(2)}_1 = 0$ was taken into account.

It can be easily seen that all minors of order two evaluate to zero if and only if $\ket{\phi}$ has all components equal to zero. Consequently, for any nonzero $\ket{\phi}\in H_E$ there are at least two linearly independent columns. According to the theory presented in Section~\ref{sec::const} on page~\pageref{Kc1}, the isometry $V$ constructed  with the Kraus operators~(\ref{Krces})  will have a completely entangled range. Acting with $V$ on  orthonormal basis states of $H_E$, we obtain the orhonormal system presented in eq.~(\ref{33ex}).

\section{\label{app:44}Details of construction of a family of $4\otimes4$ CESs of dimension $7$}

Following along the same lines as in Appendix~\ref{app:33}, we can easily come up with the operators
\begin{eqnarray}
&&K_1 = W_1\begin{pmatrix}
\sqrt{\lambda^{(1)}_1} & 0 & 0 & 0 & 0 & 0 & 0\\
0 & 0 & 0 & 0 & 0 & \sqrt{\lambda^{(1)}_6} & 0\\
0 & 0 & 0 & 0 & 0 & 0 & \sqrt{\lambda^{(1)}_7}\\
0 & 0 & 0 & \sqrt{\lambda^{(1)}_4} & 0 & 0 & 0
\end{pmatrix},\nonumber\\
&&K_2 = W_2\begin{pmatrix}
\sqrt{\lambda^{(2)}_1} & 0 & 0 & 0 & 0 & 0 & 0\\
0 & 0 & 0 & 0 & \sqrt{\lambda^{(2)}_5} & 0 & 0\\
0 & 0 & \sqrt{\lambda^{(2)}_3} & 0 & 0 & 0 & 0\\
0 & 0 & 0 & 0 & 0 & 0 & \sqrt{\lambda^{(2)}_7}
\end{pmatrix},\nonumber\\
&&K_3 = W_3\begin{pmatrix}
0 & 0 & 0 & 0 & 0 & 0 & 0\\
0 & \sqrt{\lambda^{(3)}_2} & 0 & 0 & 0 & 0 & 0\\
0 & 0 & \sqrt{\lambda^{(3)}_3} & 0 & 0 & 0 & 0\\
0 & 0 & 0 & \sqrt{\lambda^{(3)}_4} & 0 & 0 & 0
\end{pmatrix},\nonumber\\
&&K_4 = W_4\begin{pmatrix}
0 & 0 & 0 & 0 & \sqrt{\lambda^{(4)}_5} & 0 & 0\\
0 & \sqrt{\lambda^{(4)}_2} & 0 & 0 & 0 & 0 & 0\\
0 & 0 & 0 & 0 & 0 & \sqrt{\lambda^{(4)}_6} & 0\\
0 & 0 & 0 & 0 & 0 & 0 & 0
\end{pmatrix},
\end{eqnarray}
such that all $K_i^{\dagger}K_i$ are diagonal. Here $W_i$ are $4\times4$ unitary matrices. The operators $K_i$ will satisfy eq.~(\ref{TrP}) on condition that any two lambdas with the same subscript add up to $1$.

Setting $W_1=I$ and choosing $W_2$, $W_3$, $W_4$ to be permutations
\begin{equation}
    W_2\colon\,\left(\begin{array}{cc}
     1\, 3\\
     3\, 1 
\end{array}\right),\quad W_3\colon\,
\left(\begin{array}{cc}
     3\, 4\\
     4\, 3
\end{array}\right), \quad W_4\colon\,
\left(\begin{array}{cc}
     2\, 4\\
     4\, 2
\end{array}\right),
\end{equation}
we obtain proper Kraus operators~(in the described above sense). Reconstructing associated isometry via eq.~(\ref{IsoK}), we obtain the orthonormal system of vectors spanning a CES and presented in eqs.~(\ref{sfs}).

\section{\label{app:333}Details of construction of  a $3\otimes3\otimes3$ GES of maximal dimension}

The operators $P^{(i)}_{\alpha}$, defined by eqs.~(\ref{PrP})~-~(\ref{Ac}), are the constituent parts of $K_i^{\dagger}K_i$ in scheme~(\ref{Kstruct}). When we write $K_i$ themselves in the form of the singular value decomposition~(\ref{svd333}), the square roots of $P^{(i)}$ are involved, so it is convenient to consider the diagonalization of these operators:
\begin{equation}
    P^{(i)}_{\alpha} = U_i(\alpha)\,D_i(\alpha)\,U_i(\alpha)^{\dagger},\quad i=1,\,2,\,3,
\end{equation}
where it is easy to obtain that the matrices $U_i(\alpha)$ and $D_i(\alpha)$ are defined by
\begin{subequations}\label{Pdiag}
\begin{equation}
U_1(\alpha) = \begin{pmatrix}
\cos{\alpha} & -\sin{\alpha}\\
\sin{\alpha} & \cos{\alpha}
\end{pmatrix},\quad
D_1(\alpha) = \begin{pmatrix}
\lambda^{(1)}_{\alpha} & 0\\
0 & 0
\end{pmatrix},
\end{equation}
\begin{equation}
U_2(\alpha) = \begin{pmatrix}
\sin{\alpha} & -\cos{\alpha}\\
\cos{\alpha} & \sin{\alpha}
\end{pmatrix},\quad
D_2(\alpha) = \begin{pmatrix}
\lambda^{(2)}_{\alpha} & 0\\
0 & 0
\end{pmatrix},
\end{equation}
\begin{equation}
U_3(\alpha) = \frac1{\sqrt2}\begin{pmatrix}
1 & 1\\
-1 & 1
\end{pmatrix},\quad
D_3(\alpha) = \begin{pmatrix}
\lambda^{(3)}_{\alpha} & 0\\
0 & 0
\end{pmatrix},
\end{equation}
\end{subequations}
and the nonzero eigenvalues of $P^{(i)}_{\alpha}$ are
\begin{eqnarray}
\lambda^{(1)}_{\alpha} = &\lambda^{(2)}_{\alpha}& = \frac1{(1+\sin{2\alpha})},\nonumber\\
\lambda^{(3)}_{\alpha} &=& \frac{2\sin{2\alpha}}{(1+\sin{2\alpha})}.
\end{eqnarray}

Now we can write out the ingredients of eq.~(\ref{svd333}). The matrices $V_i$, $i=1,\,2,\,3$, are block-diagonal $16\times16$ unitaries
\begin{widetext}
\begin{equation}\label{Vi}
V_i = \begin{pmatrix}
U_i(\alpha_1)^{\dagger}\\
& U_i(\alpha_2)^{\dagger}\\
& & U_i(\alpha_3)^{\dagger}\\
& & & U_i(\alpha_4)^{\dagger}\\
& & & & U_i(\alpha_5)^{\dagger}\\
& & & &  & 1\\
& & & &  & & & & & 1\\
& & & &  & & & & & & & & & 1\\
& & & &  & & & & & & & & & & & & & 1\\
& & & &  & & & & & & & & & & & & & & & & & 1 \\
& & & &  & & & & & & & & & & & & & & & & & & & & & 1\\
\end{pmatrix}
\end{equation}
\end{widetext}
The matrices $\Sigma_i$ are expressed in terms of $\lambda^{(i)}_j$ and $\lambda^{(i)}_{\alpha_j}$ as
\begin{widetext}
\begin{equation}
    \Sigma_1 = \begin{pmatrix}
    \sqrt{\lambda^{(1)}_{\alpha_1}} & 0 & 0 & 0 & 0 & 0 & 0 & 0 & 0 & 0 & 0 & 0 & 0 & 0 & 0 & 0\\
    0 & 0 & 0 & 0 & 0 & 0 & 0 & 0 & 0 & 0 & \sqrt{\lambda^{(1)}_1} & 0 & 0 & 0 & 0 & 0\\
    0 & 0 & \sqrt{\lambda^{(1)}_{\alpha_2}} & 0 & 0 & 0 & 0 & 0 & 0 & 0 & 0 & 0 & 0 & 0 & 0 & 0\\
    0 & 0 & 0 & 0 & 0 & 0 & 0 & 0 & 0 & 0 & 0 & 0 & \sqrt{\lambda^{(1)}_2} & 0 & 0 & 0\\
    0 & 0 & 0 & 0 & \sqrt{\lambda^{(1)}_{\alpha_3}} & 0 & 0 & 0 & 0 & 0 & 0 & 0 & 0 & 0 & 0 & 0\\
    0 & 0 & 0 & 0 & 0 & 0 & 0 & 0 & 0 & 0 & 0 & 0 & 0 & \sqrt{\lambda^{(1)}_3} & 0 & 0\\
    0 & 0 & 0 & 0 & 0 & 0 & \sqrt{\lambda^{(1)}_{\alpha_4}} & 0 & 0 & 0 & 0 & 0 & 0 & 0 & 0 & 0\\
    0 & 0 & 0 & 0 & 0 & 0 & 0 & 0 & 0 & 0 & 0 & 0 & 0 & 0 & 0 & \sqrt{\lambda^{(1)}_4}\\
    0 & 0 & 0 & 0 & 0 & 0 & 0 & 0 & \sqrt{\lambda^{(1)}_{\alpha_5}} & 0 & 0 & 0 & 0 & 0 & 0 & 0\\
    \end{pmatrix}
\end{equation}
\begin{equation}
    \Sigma_2 = \begin{pmatrix}
    \sqrt{\lambda^{(2)}_{\alpha_1}} & 0 & 0 & 0 & 0 & 0 & 0 & 0 & 0 & 0 & 0 & 0 & 0 & 0 & 0 & 0\\
    0 & 0 & 0 & 0 & 0 & 0 & 0 & 0 & 0 & 0 & 0 & \sqrt{\lambda^{(2)}_1} & 0 & 0 & 0 & 0\\
    0 & 0 & \sqrt{\lambda^{(2)}_{\alpha_2}} & 0 & 0 & 0 & 0 & 0 & 0 & 0 & 0 & 0 & 0 & 0 & 0 & 0\\
    0 & 0 & 0 & 0 & 0 & 0 & 0 & 0 & 0 & 0 & 0 & 0 & \sqrt{\lambda^{(2)}_2} & 0 & 0 & 0\\
    0 & 0 & 0 & 0 & \sqrt{\lambda^{(2)}_{\alpha_3}} & 0 & 0 & 0 & 0 & 0 & 0 & 0 & 0 & 0 & 0 & 0\\
    0 & 0 & 0 & 0 & 0 & 0 & 0 & 0 & 0 & 0 & 0 & 0 & 0 & 0 &\sqrt{\lambda^{(2)}_3} & 0\\
    0 & 0 & 0 & 0 & 0 & 0 & \sqrt{\lambda^{(2)}_{\alpha_4}} & 0 & 0 & 0 & 0 & 0 & 0 & 0 & 0 & 0\\
    0 & 0 & 0 & 0 & 0 & 0 & 0 & 0 & 0 & 0 & 0 & 0 & 0 & 0 & 0 & \sqrt{\lambda^{(2)}_4}\\
    0 & 0 & 0 & 0 & 0 & 0 & 0 & 0 & \sqrt{\lambda^{(2)}_{\alpha_5}} & 0 & 0 & 0 & 0 & 0 & 0 & 0\\
    \end{pmatrix}
\end{equation}
\begin{equation}
    \Sigma_3 = \begin{pmatrix}
    \sqrt{\lambda^{(3)}_{\alpha_1}} & 0 & 0 & 0 & 0 & 0 & 0 & 0 & 0 & 0 & 0 & 0 & 0 & 0 & 0 & 0\\
    0 & 0 & 0 & 0 & 0 & 0 & 0 & 0 & 0 & 0 & \sqrt{\lambda^{(3)}_1} & 0 & 0 & 0 & 0 & 0\\
    0 & 0 & \sqrt{\lambda^{(3)}_{\alpha_2}} & 0 & 0 & 0 & 0 & 0 & 0 & 0 & 0 & 0 & 0 & 0 & 0 & 0\\
    0 & 0 & 0 & 0 & 0 & 0 & 0 & 0 & 0 & 0 & 0 & \sqrt{\lambda^{(3)}_2} & 0 & 0 & 0 & 0\\
    0 & 0 & 0 & 0 & \sqrt{\lambda^{(3)}_{\alpha_3}} & 0 & 0 & 0 & 0 & 0 & 0 & 0 & 0 & 0 & 0 & 0\\
    0 & 0 & 0 & 0 & 0 & 0 & 0 & 0 & 0 & 0 & 0 & 0 & 0 & \sqrt{\lambda^{(3)}_3} & 0 & 0\\
    0 & 0 & 0 & 0 & 0 & 0 & \sqrt{\lambda^{(3)}_{\alpha_4}} & 0 & 0 & 0 & 0 & 0 & 0 & 0 & 0 & 0\\
    0 & 0 & 0 & 0 & 0 & 0 & 0 & 0 & 0 & 0 & 0 & 0 & 0 & 0 & \sqrt{\lambda^{(3)}_4} & 0\\
    0 & 0 & 0 & 0 & 0 & 0 & 0 & 0 & \sqrt{\lambda^{(3)}_{\alpha_5}} & 0 & 0 & 0 & 0 & 0 & 0 & 0\\
    \end{pmatrix}
\end{equation}
\end{widetext}
The structure of each matrix $\Sigma_i^{\dagger}\Sigma_i$ is similar to the one in eq.~(\ref{Kstruct}) with the only distinction that all blocks $P^{(i)}_{\alpha_j}$ are in the diagonal forms $D_i(\alpha_j)$ given by eq.~(\ref{Pdiag}). Multiplication of $\Sigma_i^{\dagger}\Sigma_i$ by $V_i$ from eq.~(\ref{Vi}) recovers the exact form of $K_i^{\dagger}K_i$ presented in eq.~(\ref{Kstruct}). The unitaries $W_i$ from eq.~(\ref{svd333}) don't change this form and can be chosen arbitrarily for our purpose.

Our particular choice of values for the variables $\lambda^{(i)}_j$ and $\alpha_j$ in scheme~(\ref{Kstruct}) is
\begin{eqnarray}\label{InSet}
\alpha_i = \alpha \equiv \pi/6,\quad i=1,\ldots5,&&\nonumber\\
\lambda^{(1)}_1 = \frac23,\qquad \lambda^{(3)}_1 = \frac13,&&\nonumber\\
\lambda^{(2)}_1 = \frac14,\qquad \lambda^{(3)}_2 = \frac34,&&\nonumber\\
\lambda^{(1)}_2 = \frac16,\qquad \lambda^{(2)}_2 = \frac56,&&\nonumber\\
\lambda^{(1)}_3 = \frac13,\qquad \lambda^{(3)}_3 = \frac23,&&\nonumber\\
\lambda^{(2)}_3 = \frac34,\qquad \lambda^{(3)}_4 = \frac14,&&\nonumber\\
\lambda^{(1)}_4 = \frac56,\qquad \lambda^{(2)}_4 = \frac16.&&
\end{eqnarray}
We set all $\alpha_i$ equal to each other for simplicity.

To satisfy the conditions on $K_i$ described in Section~\ref{sec::const} on page~\pageref{Kc1}, we search for the proper unitaries $W_i$ in the form of permutation matrices. Using the procedure described in Section~\ref{sec::const} on page~\pageref{svd333} and the Groebner basis algorithm, we came up with the three proper permutations. Let $e_i$ denote a $9$-dimensional vector with the i-th component equal to $1$ and the rest equal to zero. Such vectors can serve as columns of a permutation matrix. The unitaries $W_i$ are given by
\begin{eqnarray}\label{WP}
W_1 &=& \left(e_1\,e_2\,e_3\,e_4\,e_7\,e_8\,e_5\,e_6\,e_9\right)\nonumber\\
W_2 &=& \left(e_6\,e_4\,e_7\,e_5\,e_8\,e_1\,e_9\,e_3\,e_2\right)\nonumber\\
W_3 &=& \left(e_4\,e_7\,e_8\,e_1\,e_3\,e_6\,e_2\,e_9\,e_5\right).
\end{eqnarray}

Using eqs.~(\ref{IsoK}), (\ref{svd333}), (\ref{Pdiag})-(\ref{WP}), we obtain an orthonormal system of vectors spanning a $9\otimes3$ CES:
\begin{widetext}
\begin{eqnarray}\label{Sp333}
&&\ket{\psi}_{1,2}=c_{1,2}\ket{0}\otimes\ket{0} + c_{2,1}\ket{5}\otimes\ket{1}\, \pm\,c_3\ket{3}\otimes\ket{2},\quad \ket{\psi}_{3,4}=\pm c_3\ket{2}\otimes\ket{0} + c_{1,2}\ket{6}\otimes\ket{1}
+ c_{2,1}\ket{7}\otimes\ket{2},\nonumber\\
&&\ket{\psi}_{5,6}=c_{2,1}\ket{6}\otimes\ket{0}\,\pm\, c_3\ket{7}\otimes\ket{1}
+c_{1,2}\ket{2}\otimes\ket{2},\quad \ket{\psi}_{7,8}=c_{1,2}\ket{4}\otimes\ket{0}\,\pm\, c_3\ket{8}\otimes\ket{1}
+c_{2,1}\ket{1}\otimes\ket{2},\nonumber\\
&&\ket{\psi}_{9,10}=\pm c_3\ket{8}\otimes\ket{0} + c_{2,1}\ket{1}\otimes\ket{1}
+ c_{1,2}\ket{4}\otimes\ket{2},\quad \ket{\psi}_{11}=\sqrt{\lambda^{(1)}_1}\ket{1}\otimes\ket{0} + \sqrt{\lambda^{(3)}_1}\ket{6}\otimes\ket{2},\nonumber\\
&&\ket{\psi}_{12}=\sqrt{\lambda^{(2)}_1}\ket{3}\otimes\ket{1} + \sqrt{\lambda^{(3)}_2}\ket{0}\otimes\ket{2},\: \ket{\psi}_{13}=\sqrt{\lambda^{(1)}_2}\ket{3}\otimes\ket{0} + \sqrt{\lambda^{(2)}_2}\ket{4}\otimes\ket{1},\:\ket{\psi}_{14}=\sqrt{\lambda^{(1)}_3}\ket{7}\otimes\ket{0}\nonumber\\ &&+ \sqrt{\lambda^{(3)}_3}\ket{5}\otimes\ket{2},
\ket{\psi}_{15}=\sqrt{\lambda^{(2)}_3}\ket{0}\otimes\ket{1} + \sqrt{\lambda^{(3)}_4}\ket{8}\otimes\ket{2},\quad \ket{\psi}_{16}=\sqrt{\lambda^{(1)}_4}\ket{5}\otimes\ket{0} + \sqrt{\lambda^{(2)}_4}\ket{2}\otimes\ket{1},\nonumber\\
&&c_1 = \frac{\cos{\alpha}}{\sqrt{1+\sin{2\alpha}}},\qquad c_2 = \frac{\sin{\alpha}}{\sqrt{1+\sin{2\alpha}}} ,\qquad c_3 = \sqrt{\frac{\sin{2\alpha}}{1+\sin{2\alpha}}},
\end{eqnarray}
\end{widetext}
where the subscript indices separated by comma along with the `$\pm$' sign imply that the first index is used simultaneously with the `$+$' sign and the second index  with the `$-$' sign in the corresponding expression.  

Next, we introduce the following scheme for the factorization of the $9$-dimensional subsystem into two $3$-dimensional ones:
\begin{eqnarray}\label{sch3}
\ket{0}\qquad&\rightarrow&\qquad\ket{2}\otimes\ket{2},\nonumber\\
\ket{1}\qquad&\rightarrow&\qquad\ket{2}\otimes\ket{0},\nonumber\\
\ket{2}\qquad&\rightarrow&\qquad\ket{2}\otimes\ket{1},\nonumber\\
\ket{3}\qquad&\rightarrow&\qquad\ket{1}\otimes\ket{0},\nonumber\\
\ket{4}\qquad&\rightarrow&\qquad\ket{0}\otimes\ket{2},\nonumber\\
\ket{5}\qquad&\rightarrow&\qquad\ket{0}\otimes\ket{1},\nonumber\\
\ket{6}\qquad&\rightarrow&\qquad\ket{0}\otimes\ket{0},\nonumber\\
\ket{7}\qquad&\rightarrow&\qquad\ket{1}\otimes\ket{2},\nonumber\\
\ket{8}\qquad&\rightarrow&\qquad\ket{1}\otimes\ket{1}.
\end{eqnarray}
We've made such a choice trying to transform the vectors $\ket{\psi}_1\,\ldots,\,\ket{\psi}_{10}$ of eq.~(\ref{Sp333}) to entangled ones. At the same time, the vectors $\ket{\psi}_{11}$ and $\ket{\psi}_{16}$ are transformed to separable ones~(the basis vectors $\ket{2}$, $\ket{5}$, $\ket{6}$ of the $9$-dimensional subsystem are involved), and we couldn't come up with a better scheme leaving all these vectors entangled. To come around this problem, we return to the CES defined by eq.~(\ref{Sp333}) and mix the  vectors $\ket{2}$, $\ket{5}$, $\ket{6}$  with each other  by means of a simple $3\times3$ unitary transformation:
\begin{eqnarray}\label{U3}
\ket{2}&\rightarrow -&\frac13\ket{2} + \frac23\ket{5} + \frac23\ket{6},\nonumber\\
\ket{5}&\rightarrow& \frac23\ket{2} - \frac13\ket{5} + \frac23\ket{6}\nonumber\\
\ket{6}&\rightarrow& \frac23\ket{2} + \frac23\ket{5} - \frac13\ket{6},
\end{eqnarray}
which we took from Ref.~\cite{Tremain2011}.
Being a local unitary operation, such a substitution in eq.~(\ref{Sp333}) doesn't change entanglement of the CES. After that we proceed to the factorization~(\ref{sch3}). The analysis of the newly obtained subspace with the use of the Groebner basis algorithm shows that it is a GES.

%\nocite{*}

\bibliography{apssamp}% Produces the bibliography via BibTeX.

\end{document}